\documentclass[journal, twocolumn, twoside, 10pt] {IEEEtran}

\usepackage[section]{placeins}
\usepackage{amssymb}
\usepackage{supertabular}
\usepackage{subfigure}
\usepackage{color}
\usepackage{amsmath,bm,epsfig}
\usepackage{epstopdf}
\usepackage{float}
\usepackage{subfig}
\usepackage{array}
\usepackage{multirow}
\usepackage{graphicx}
\usepackage{enumerate}
\usepackage{soul}     % for strike-out fonts
\usepackage[normalem]{ulem}     % for strike-out fonts

\def\sec{Section}

\begin{document}

\newcolumntype{P}[1]{>{\centering\arraybackslash}p{#1}} % table columns with specified width and centered content
\renewcommand\arraystretch{1.15} % increase row height for complexity symbols and better table readability
	
\title{Automatic synchronization of multi-user photo galleries}
\author{
	E. Sansone, \and
	K. Apostolidis, \and
	N. Conci, \and
	G. Boato, \and
	V. Mezaris, \and
	F.G.B. De Natale
	
%\thanks{Manuscript received XX XX XXXX. Copyright (c) 2016 IEEE. Personal use of this material is permitted. However, permission to use this material for any other purposes must be obtained from the IEEE by sending a request to pubs-permissions@ieee.org.
%}

%\IEEEcompsocitemizethanks{
%\IEEEcompsocthanksitem{Emanuele Sansone, Nicola Conci, Giulia Boato, and Francesco G.B. De Natale are with the Department of Information Engineering and Computer Science - DISI, University of Trento, 38123, Italy; e-mail: \{e.sansone, conci, boato, denatale\}@unitn.it.}
%\IEEEcompsocthanksitem{Konstantinos Apostolidis and Vasileios Mezaris are with the Information Technologies Institute (ITI), Centre for Research and Technology Hellas (CERTH), Thermi 57001, Greece; e-mail: \{kapost, bmezaris\}@iti.gr.}
%}
}
\maketitle

% MRF approach symbols
\newtheorem{synchronization}{Definition}
\newtheorem{assumption}{Assumption}

% CERTH approach symbols
\newcommand{\geoFactor}{a_{geo}}

\begin{abstract}
In this paper we address the issue of photo galleries synchronization, where pictures related to the same event are collected by different users. Existing solutions to address the problem are usually based on unrealistic assumptions, like time consistency across photo galleries, and often heavily rely on heuristics, limiting therefore the applicability to real-world scenarios. We propose a solution that achieves better generali\-zation performance for the synchronization task compared to the available literature. The method is characterized by three stages: at first, deep convolutional neural network features are used to assess the visual similarity among the photos; then, pairs of similar photos are detected across different galleries and used to construct a graph; eventually, a probabilistic graphical model is used to estimate the temporal offset of each pair of galleries, by traversing the minimum spanning tree extracted from this graph. The experimental evaluation is conducted on four publicly available datasets covering different types of events, demonstrating the strength of our proposed method. A thorough discussion of the obtained results is provided for a critical assessment of the quality in synchronization.
\end{abstract}


\section{Introduction}
\label{sec:intro}
The proliferation of multimedia capturing devices and ubi\-quitous connectivity have increased the possibility of sharing photos and other multimedia content over the Internet through social networks, shared spaces, cloud storage systems, and various other multimedia sharing services. The design of methods and systems to efficiently manage and organize this ever-increasing amount of data represents a great challenge for the multimedia research community \cite{survey}.

\begin{figure}[ht]
	\centering
	%\captionsetup{justification=centering}
	\includegraphics[width=\linewidth]{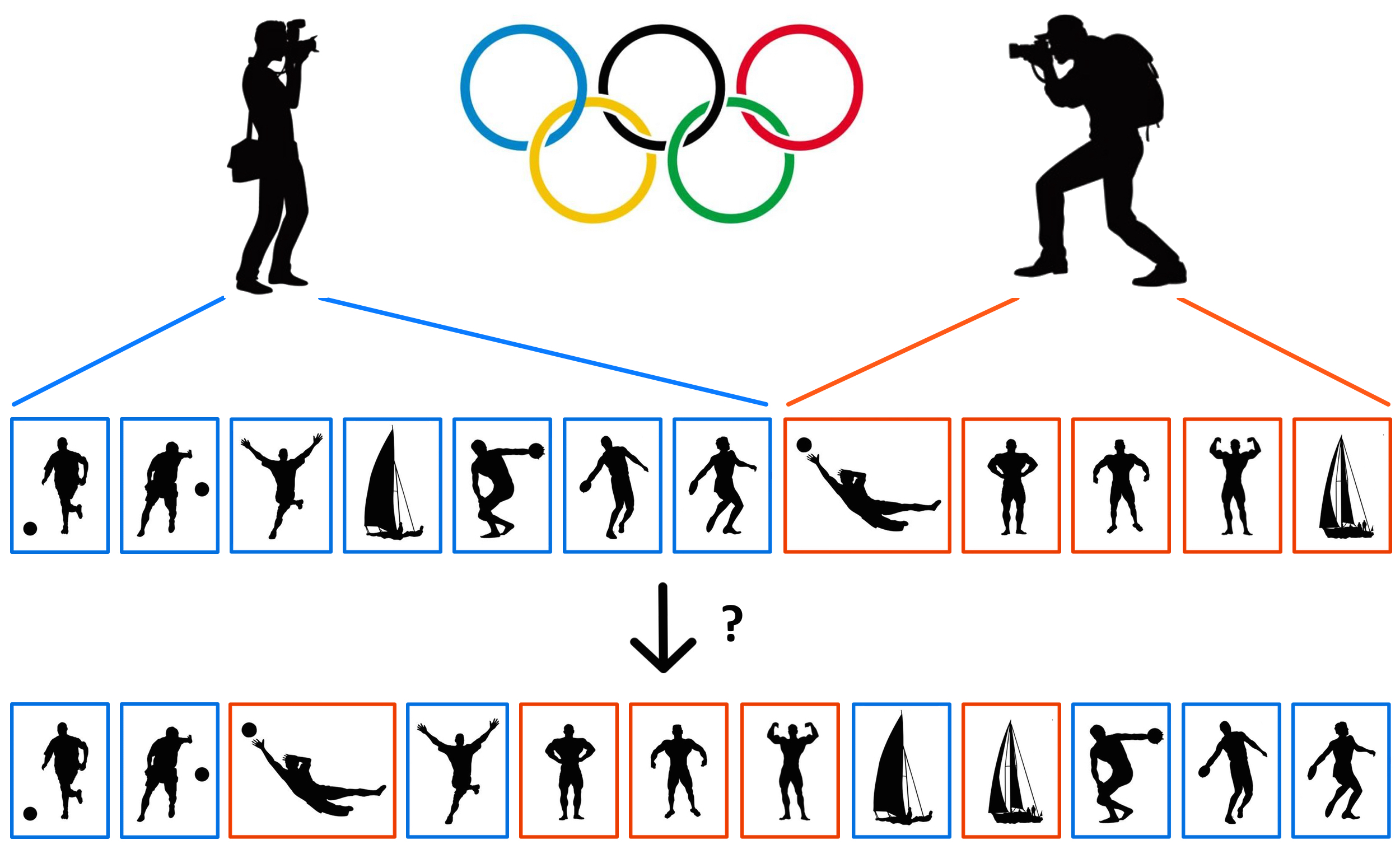}
	\caption{Illustration of the problem of synchronizing multi-user photo galleries.}
	\label{fig:problem}
\end{figure}

A situation that frequently occurs in this context is the need of users to share media captured at an event they attended (a concert, a ceremony, a sport or public event, etc.) with others and/or within a social community interested in that event. The final goal of sharing would be the creation of a shared repository, where all the contributors, and possibly users who did not attend in person, can find an enriched view of the event by observing it from multiple perspectives, with finer granularity and better completeness. Data could then be exploited in various contexts including, for example, the creation of storyboards, automatic summaries, and persona\-lized albums. An illustration of the problem is depicted in Fig. \ref{fig:problem}. However, in order to achieve this, data has to be organized consistently, so as to allow browsing across different dimensions, e.g. sorting based on the time of acquisition, clustering by location, and/or ranking by visual similarity according to a given query. 

Several approaches dealing with different facets of the gene\-ral problem of multimedia data organization have appeared in the literature. These include techniques for clustering, summarization, visualization, and semantic analysis. Clustering techniques, e.g. \cite{clust1}, aim at partitioning a collection of multimedia data into clusters of similar or related items. Summarization techniques, on the other hand, e.g. \cite{sum1,sum2}, aim at producing a concise yet informative version of the original dataset. Visuali\-zation methods, e.g. \cite{vis1}, seek to present multimedia data, be it original collections or clustered/summarized versions of them, for an enhanced user experience. Lastly, semantic analysis and annotation methods have seen great progress in the last few years. In fact, thanks to a better knowledge about \textit{how} humans perceive and interpret the multimedia data (e.g., by means of event detection \cite{event1,vid2,rosani}, person detection and identification \cite{person1,person2}, photo tagging \cite{refine1,ann4}), they provide valuable support to data organization according to a human-centered perspective. 

In order to ensure an efficient organization of media galleries, and in particular when dealing with \textit{multi-user} collections, the temporal dimension is very critical. {In such scenarios each gallery contains photos from a single user/device, therefore the timestamps of photos within a gallery are considered to be consistent, while the temporal relation of photos from different galleries is affected by an unknown offset}. A consistent time annotation allows easily aligning photos belonging to different galleries along a common axis, thus making it possible to locate and cluster important sub-events, match different users' views of the same happening, and ultimately create a unique and consistent flow of information, the \textit{story}. Therefore, the lack of temporal data annotation or its limited accuracy/reliability may cause major problems in media management. This is indeed the most common situation: although we can reasonably assume the coherence of the timestamps provided by a single device within the time-frame of an event, nothing or little can be said about the absolute correctness of such time-based annotations, and thus we cannot guarantee the coherence of timestamps generated by different devices. Possible sources of error include wrong or missing setting of the camera clock, the use of multimedia capturing devices with different time zone settings, missing timestamp information, post-processing of the media with modification or removal of timestamps. These situations prevent a correct synchronization of the different galleries provided by different users, which is indeed necessary for an effective organization of the data.

In this scenario it is necessary to estimate the temporal offset among the clocks of the different users' devices, taking into account all the available information: the metadata of the photos, when available, often stored in the EXIF header, together with the visual content of the images. As far as the EXIF metadata is concerned, we consider date and time of acquisition, together with the geo-location, according to the information provided by the GPS receiver. Note that the availability of such information is not a pre-requisite for synchronization, and in particular the GPS coordinates are often missing, while timestamps are considered consistent only for the photos captured by the same device (referred to, as a gallery).

In this paper, we present a method to achieve an automatic synchronization of multi-user galleries, which outperforms the methods found in the literature. Photo similarity is assessed by extracting Deep Convolutional Neural Network (DCNN) features. Similar photos are then used to define a graph of galleries. Each pair of galleries in the graph is finally synchronized using a probabilistic graphical model. %\footnote{The link to the project website with the source code will be made available upon acceptance.}.
Experimental results over different event types show that the proposed method is able to achieve very good synchronization precision and accuracy. 

The paper is structured as follows: \sec~\ref{sec:soa} reviews the related work in this research area.  \sec~\ref{sec:proposed} presents the method we propose. \sec~\ref{sec:results} reports the results of experimental evaluation and comparisons on four benchmark datasets, also including a thorough commentary on the complexity of the proposed method. Concluding remarks are provided in \sec~\ref{sec:conclusions}.

\section{Related Work}
\label{sec:soa}
Different pieces of information can be used by a single user to properly arrange his photo galleries, and the temporal data is generally ranked among the most useful elements, as demonstrated in \cite {graham2002time, mulhem2003home}. In \cite{sinha2009personal,rabbath2011automatic, pigeau2003spatio} the timestamps provided in the metadata of digital photos are used for photo gallery organization purposes, as for example alignment and summarization. Such approaches are shown to be very effective on galleries captured by a single device, where the provided timestamps are consistent. 
Extending the photo organization use case to multiple cameras (as in the case of shared repositories of photo galleries, where different users share their pictures and videos of the same event), the overall consistency of the timestamps becomes questionable. This is due to the fact that different users may have different time settings (time zones, wrong clock setting) in their devices, and, as a consequence, directly using these timestamps can introduce noise in any photo organization process.

The literature reports some works that, although not matching exactly the problem we are addressing, do deal with multi-user photo galleries and introduce some notion of information alignment. In \cite{kim2014reconstructing}, storyline graphs are produced from web communities' photos, for supporting photo recommendation applications. In \cite{kim2014joint}, still photos are temporally aligned with video footage, while in \cite{kim2015joint} photos of different blog posts are grouped using natural language processing of the text found in the blogs. The above methods focus on summarizing media collections and, when performing media alignment with the use of photo timestamps, they treat them as if they were consistent across different photo galleries. A different photo-sequencing problem is addressed in \cite{moses2013space}: given a set of images depicting the same object, temporal ordering is achieved by computing the optimal similarity transformation of all images to the first image in the sequence. The warped images can then be animated in the computed temporal order, as a single video sequence. However, this method makes the strong assumption that all pictures depict the same rigid object. Looking a bit beyond images, the alignment of different time-series, which are typically obtained by repeated measurements of a biological, chemical or physical process, is a well studied problem (e.g. the method in \cite{suematsu2012time}). However, such methods rely on finding common (exact-duplicate) sub-sequences between two data time-series, and their application to our temporal photo alignment problem is therefore not straightforward.

\begin{figure}[t!]
	\centering
	%\captionsetup{justification=centering}
	\includegraphics[width=0.8\linewidth]{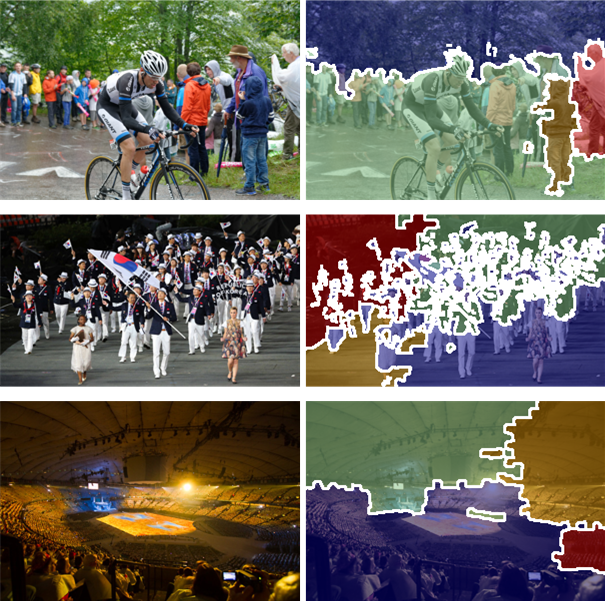}
	\caption{Examples of the segmentation algorithm of \cite{mfc} applied on a selection of photos from the datasets we use for our experiments.}
	\label{fig:mfc}
\end{figure}

Previous methods that deal with our problem include \cite{kim2013jointly,broilo:icip,yang2012photo,oliva2001modeling, wang2010locality}. In \cite{kim2013jointly}, a collective storyline of photos is constructed by exploiting the visual information. Each image is segmented in foreground objects and background, so as to assess photo similarity by detecting instances of the same objects across galleries. A user-defined parameter \textit{K} denotes the number of foreground areas in which a photo can be split, using a Multiple Foreground Co-segmentation algorithm \cite{mfc}. However, real-world photo collections are not always centered on objects that can be easily segmented, e.g., photos of a music festival may contain faces from the audience, as well as musicians faces along with a complex background. An example is provided in Fig. \ref{fig:mfc}, where we show that the segmentation results obtained by applying the algorithm proposed in \cite{mfc} for a selection of photos in the real-life datasets used in this paper. From these examples it is clear that this approach is not suitable for our synchronization scenario, due to the imperfections of this (and any other) segmentation algorithm. Broilo et al. \cite{broilo:icip} attempt a content-based alignment approach to compute the estimated delay among the photos taken with different cameras by finding the suitable pairs of similar photos across the different galleries of the same event; the delay among the galleries is estimated based on the time delays between the selected pairs of pictures. In order to operate properly, however, the method relies on the hypothesis that different photographers take photos that capture the same sub-events. This condition turns out to be very restrictive, especially when considering a real and unsupervised scenario, where no control on the acquisition process can be imposed. Along with visual data, geo-location information can also be used for alignment purposes, such as in the approach by Yan et al. \cite{yang2012photo}, in which a bipartite kernel sparse representation of visual and spatial similarities is used for photo stream alignment and summarization. However, the framework of \cite{yang2012photo} requires considerable computational power and cannot scale well beyond approximately a hundred images.
Several recent works that study the synchronization problem of generic multi-user photo galleries have been proposed and evaluated in relation to the MediaEval SEM (Synchronization of multi-user Event Media) benchmarking activity that was organized in 2014 \cite{conci:sem} and 2015 \cite{sem15_overview}. Concerning the features used for similarity assessment of the images, Zaharieva et al. \cite{zaharieva2014multimodal} propose exploiting the MPEG-7 Color Structure Descriptor \cite{messing2001mpeg} and a Joint Composite Descriptor. In \cite{apostolidis2014certh,semJRS,sem15jrs} the SIFT local descriptor is used, while in \cite{semMRF} the SURF local descriptor and HSV color histograms are employed. 
%As far as the estimation of the offsets is concerned, an Agglomerative Hierarchical Clustering method is used in \cite{zaharieva2014multimodal} that exploits the similarity of the photos to produce a hierarchy of clusters. The lowest-level clusters are considered as very similar photos between different galleries and are used to compute the temporal offsets between different galleries. The highest-level clusters are considered as sub-events. 
In \cite{apostolidis2014certh}, the most similar photos between different galleries are detected and a graph of photo similarities is employed to find paths between each gallery and the reference one. In \cite{semJRS}, temporal offsets are expressed as a non-homogeneous linear equation system, and an approximate solution is calculated. In \cite{semMRF}, a probabilistic graphical model is built, in which each temporal displacement is identified by a set of nearest-neighbor photo pairs across the galleries. The method in \cite{sem15certh} is an extension of the method in \cite{apostolidis2014certh} that can also handle galleries that include video and audio. Finally, in \cite{sem15jrs} a probabilistic algorithm is employed, where in each run a hypothesis is calculated for the set of time offsets with respect to a reference gallery. The final set of time offsets is calculated as the medoid of all hypotheses. The most relevant works mentioned in this section are summarized in Table~\ref{tab:bib_review}.

{
In this work we build on our preliminary methods \cite{huevent15apo} and \cite{semMRF}, which we combine and extend in several ways. Specifically, the novelties of the proposed method compared to our older methods (or a straightforward combination thereof), and to the methods of the literature are: \begin{enumerate}[i)]
	\item The use of DCNN-based features to assess the semantic similarity of photos. By replacing the hand-crafted features of \cite{huevent15apo}, \cite{semMRF} with DCNN-based features, we take advantage of the significant progress made in deep learning. To our knowledge, this is the first work that uses this kind of features for the specific problem of multi-user photo gallery synchronization.
	\item The combination of graph-based representation of galleries and a probabilistic model to perform the synchronization. Graph-based representations in image and video analysis problems are not new (e.g. in \cite{new_lin_shot} and \cite{new_lin_scene} the problems of video-shot and image scene categorization, respectively, are formed as graph partitioning tasks; similarly, the photo synchronization problem in \cite{huevent15apo} is transformed to a graph-search). Also, the use of a probabilistic model for photo synchronization was first briefly sketched in \cite{semMRF}. However, by extending and combining these two notions we obtain significant benefits: a) we remove the constraint of a fixed reference gallery, by employing the probabilistic model on top of the constructed graph of galleries (in contrast to \cite{semMRF}, which made the restrictive assumption that a single reference gallery was specified in advance, and all other galleries shared similar photos with it). b) We relieve the probabilistic graphical model from the need of using visual features, which are only employed at the graph construction stage to assess the visual similarity between photos and to define the potential temporal offsets. c) Compared to \cite{semMRF}, in which all possible offsets are taken into account, we treat as potential candidates in the probabilistic graphical model only the finite number of values that correspond to the time differences between similar photos across each pair of galleries in the graph. Overall, the changes to the model lead to a speed-up of three orders of magnitude in the inference stage, compared to \cite{semMRF}. d) Adopting the aforementioned probabilistic model, we perform global optimization on the offsets available for each pair of galleries (as opposed to \cite{huevent15apo}).
     \item The construction of an overall photo synchronization method that is free of thresholds (as opposed to both \cite{semMRF} and \cite{huevent15apo}) and thus is able to yield good results in diverse datasets.
\end{enumerate}} 
These elements guarantee the {effectiveness} of the proposed approach on all datasets, as will be discussed in Section \ref{sec:results}.

\section{Temporal Synchronization of Multi-User Photo Galleries}
\label{sec:proposed}
\subsection{{Problem Statement}}
\label{sec:problem}
\begin{figure}[t]
	\centering
	%\captionsetup{justification=centering}
	\includegraphics[trim={2in 2in 2in 3in},width=0.9\linewidth]{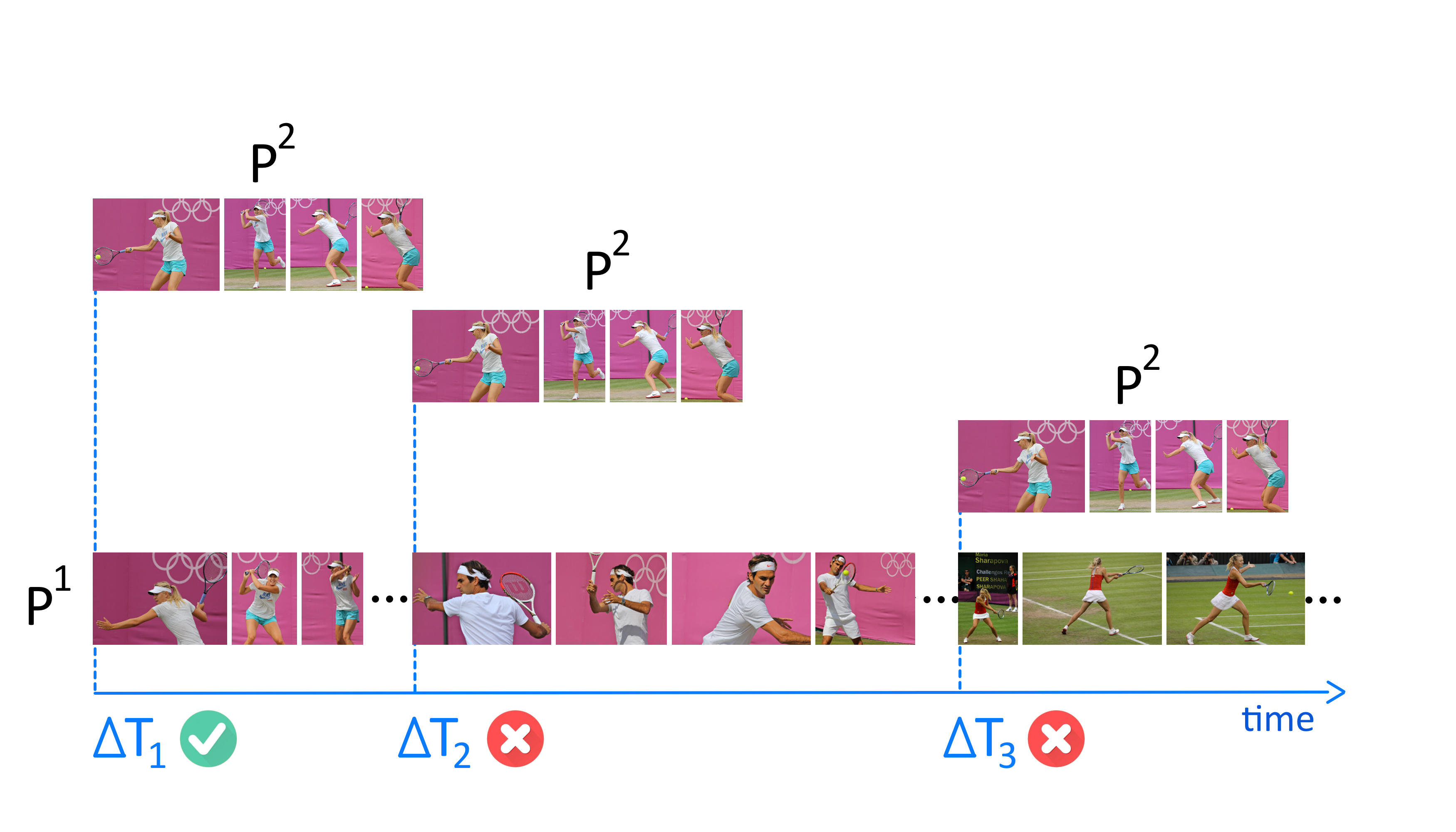}
	\caption{Example of galleries taken from one of the datasets used in this work. On the bottom, the reference gallery. On the top, the second gallery that has to be synchronized with respect to the reference according to three possible hypotheses (i.e., three possible offsets). In this case, $\Delta T^*=\Delta T_1$.}
	\label{fig:test}
\end{figure}
Let us assume we are given two photo galleries, namely $P^1=(I_1^1,I_2^1,\dots,I_{N_1}^1)$ and $P^2=(I_1^2,I_2^2,\dots,I_{N_2}^2)$, with the associated sequence of timestamps $T^1=(t_1^1,t_2^1,\dots,t_{N_1}^1)$ and $T^2=(t_1^2,t_2^2,\dots,t_{N_2}^2)$. In this case, $I_l^i$ is the $l$-th photo of the $i$-th gallery and $t_l^i$ is the corresponding timestamp. Considering that each gallery is acquired by a single device, we can assume that the relative time differences between the elements in $P^1$ (and similarly for $P^2$) are correct. 
%Conversely we cannot be sure about the relative time alignment between $T^1$ and $T^2$, since it may be affected by an unknown offset. 
Conversely, we do not know if the two sets $T^1$ and $T^2$ have the same absolute reference time or if they have been affected by an unknown offset. Therefore the problem consists of estimating the value of this temporal offset, called $\Delta T^*$, {and estimating one (in general, different) such value for each different pair of galleries.} A visual representation of the problem is shown in Fig. \ref{fig:test}.
%
%From this definition that synchronizing two photo galleries corresponds to estimating the unknown temporal offset between them.
%We then define two galleries as synchronized when the difference between the estimated offset and the ground truth offset is lower than a threshold.
%
%
\subsection{General Approach}
\label{sec:overview}
\setlength\extrarowheight{1pt}
\begin{table*}[!ht]
  \footnotesize
  \centering
\caption{Review of the most relevant works presented in \sec~\ref{sec:soa}. For each reference, the objective, the features used and the datasets adopted for validation are reported. The indication \textit{personal collection} refers to non-standard datasets retrieved from the Internet or from private material. The last column reports the main limitation of each approach.}
  \begin{tabular}{|P{0.07\textwidth} | P{0.10\textwidth} | P{0.17\textwidth} | P{0.15\textwidth}|  P{0.38\textwidth}|}
    \hline 
    \textbf{Refs.} & \textbf{Objective} & \textbf{Features} & \textbf{Dataset} & \textbf{Notes} \\ \hline 
    \cite{sinha2009personal}\cite{rabbath2011automatic}\cite{pigeau2003spatio} & Summarization  & Visual, Metadata, Text & CeWe 2009, Collection from Facebook & Galleries from a single camera, timestamps assumed reliable \\ \hline 
    \cite{kim2014reconstructing}\cite{kim2014joint}\cite{kim2015joint} & Alignment, summarization & Visual, Text & Collection from Flickr, YouTube & Timestamps assumed reliable \\ \hline 
    \cite{moses2013space} & Temporal ordering & Visual & Personal collection & All images need to depict the exact same object from different viewpoints \\ \hline 
    \cite{kim2013jointly} & Synchronization & Visual & Collection from Flickr & The same objects, which can be easily segmented, need to be depicted in the different galleries \\ \hline 
    \cite{broilo:icip} & Synchronization & Visual & Personal collection & The same sub-events must be captured in the different galleries \\ \hline 
    \cite{yang2012photo} & Synchronization & Visual + GPS & Personal collection & Difficult to scale due to high computational complexity\\ \hline 
    \cite{zaharieva2014multimodal} & Synchronization & MPEG-7 CSD & MediaEval2014 & Synchronization is based on pairwise comparison of galleries (limited scalability with respect to the number of galleries) \\ \hline
    \cite{semJRS}& Synchronization & Visual (SIFT) & MediaEval2014 & Synchronization is based on pairwise comparison of galleries (limited scalability with respect to the number of galleries) \\ \hline
    \cite{apostolidis2014certh} & Synchronization & Visual (SIFT) & MediaEval2014 & - \\ \hline
    \cite{semMRF} & Synchronization & Visual (SURF+HSV) &  MediaEval2014 & A reference gallery that covers most of the event must be provided \\ \hline 
    \cite{huevent15apo} & Synchronization & Visual (SIFT+GIST+HSV) & MediaEval2014 & -  \\ \hline
    \cite{sem15jrs}\cite{sem15certh} & Synchronization & Visual + Audio & MediaEval2015 & -  \\ 
    \hline 
  \end{tabular} 
  \label{tab:bib_review}
\end{table*}

\begin{figure}[b!]
	\centering
	%\captionsetup{justification=centering}
	\includegraphics[width=0.95\linewidth]{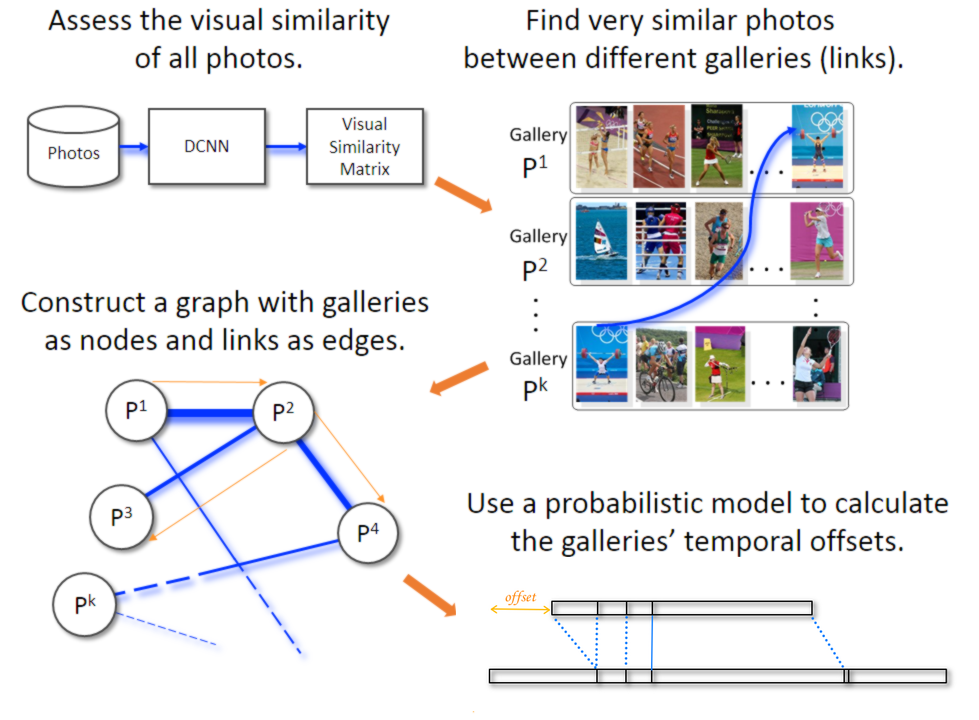}
	\caption{Overview of the proposed method.}
	\label{fig:overview}
\end{figure}

Figure \ref{fig:overview} illustrates the main stages of the proposed approach for the automatic temporal synchronization of multi-user photo galleries. Initially, similar photos across different galleries are identified. Visual similarity is evaluated exploiting the features extracted from a pre-trained Deep Convolutional Neural Network (DCNN), as explained in Section~\ref{sec:similarity}. The most similar photos from different galleries are considered as links between different pairs of photo galleries. Subsequently, we construct a graph where nodes represent galleries, and edges represent the discovered links between them, as explained in Section \ref{sec:graph}. Temporal synchronization of the galleries is achieved exploiting a probabilistic graphical model. As explained in Section \ref{sec:mrf}, for this purpose visual similarity is modeled through potential functions, and the max-sum algorithm is used for the temporal offsets estimation. In order to get a fast response without sacrificing the accuracy, we apply a coarse-to-fine optimization approach.

\subsection{Photo Similarity Assessment}
\label{sec:similarity}
As can be ascertained from Section \ref{sec:soa}, local descriptors (e.g. SIFT, SURF) are most often used to produce a representation of the visual content of the photos. Driven by the success of DCNNs in large scale image classification tasks, DCNN-based visual representations are quickly gaining ground and are used in image retrieval tasks as well, exhibiting state-of-the-art performance (\cite{zagoruyko2015learning, ng2015exploiting, xie2015image, new_lin_deep_hash, song2015rank, song2015top}).

To identify similar photos of different galleries, we adopted the method proposed in \cite{ng2015exploiting}, which uses DCNN-based features to assess the similarity of photos. We choose to use the GoogLeNet network \cite{googlenet}, due to its fast response in the testing phase. The network is pre-trained on ImageNet data using the Caffe deep learning framework \cite{caffe}, and is publicly available on the Caffe model zoo\footnote{https://github.com/BVLC/caffe/wiki/Model-Zoo}. In the 22-layer deep GoogLeNet network, there are 9 inception layers sequentially connected. 
{According to \cite{googlenet}, any input image is resized to $224 \times 224$ pixels and segmented to equal and non-overlapping regions. An inception layer extracts convolutional filters’ responses from each such region of the input image, which can be used as features to describe the visual content of the specific region (the number of regions that an image is segmented to, as well as the size of each region and of the corresponding feature vector, are reported in Table~\ref{tab:conv_sizes}). In this way we generate a single collection of L2-normalized filter responses from a selected DCNN layer (i.e., one collection for all regions of all images, and all filters of that layer), and we perform k-means clustering in this collection to obtain a vocabulary of 256 words. Then, each filter response vector is assigned to its nearest visual word, enabling the subsequent encoding of all the filter responses from all regions into one feature vector using VLAD encoding \cite{vlad}. Finally, the VLAD descriptors are normalized by intra-normalization \cite{vlad_norms}.}

Taking into consideration the evaluation carried out in \cite{ng2015exploiting}, we choose to experiment with features from the ``conv2/norm2'', ``inception3a/output'', ``inception4a/output'' and  ``inception5a/output'' layers as these layers scored the best results in different datasets. For the sake of completeness we also conducted two additional experiments: we used directly the final classifier layer (``loss3/classifier'', without any further encoding of its output) and we also implemented a fusion approach on features of all the above selected layers (except the ``loss3/classifier''), obtained by averaging the corresponding similarity matrices. {The resulting feature vector size after VLAD aggregation, for the different layers we test, is reported in Table \ref{tab:conv_sizes}}. The preliminary experiments with these different layers, which lead to the final choice adopted in our method, are reported in Section~\ref{sec:comparison_sims}.

\begin{table}[tb]
	\centering
	\caption{{Feature sizes using different inception layers and the final classifier layer of GoogLeNet, according to \cite{googlenet}, and the resulting feature vector size following VLAD aggregation for the inception layers' features.}}
	\label{tab:conv_sizes}
	\begin{tabular}{|P{2.0cm}|P{1.0cm}|P{1.2cm}|P{0.9cm}|P{1.6cm}|}
		\hline
		GoogLeNet Layer Name & Number of regions & Region Size & Layer feature size & Resulting feature vector size using a vocabulary of 256 centers\\ \hline \hline
		conv2/norm2 	   & 28 x 28 & 8 x 8	& 192 & 49152 \\ \hline
		inception3a/output & 28 x 28 & 8 x 8	& 256 & 65536 \\ \hline
		inception4a/output & 14 x 14 & 16 x 16	& 512 & 131072 \\ \hline
        inception5a/output & 7 x 7   & 32 x 32	& 832 & 212992 \\ \hline
        loss3/classifier   & 1   	 & 224 x 224	& 1000 & N/A \\ \hline
	\end{tabular}
\end{table}

We construct the visual similarity matrix $\bm{W}$ of all images in a collection as:
\begin{equation}
\bm{W}(i,j) = \exp\{-D(V_i,V_j)\},\:\forall i,j \in \{1,\dots,N\}
\end{equation}
where $V_i$ and $V_j$ are the VLAD vectors of $I_i$ and $I_j$ photos, respectively, $D$ is the Euclidean distance function and $N$ is the total number of images in the collection. Each pair of photos $I_i^k$ and $I_j^l$, where $k \neq l$ (i.e., the photos do not belong to the same gallery), is treated as a potential link between the $k$ and $l$ galleries. To define the number of links that we utilize for each pair of galleries we introduce two alternative approaches:
\begin{itemize}
	\item \textit{exact} approach: for each pair of galleries we keep the $\left \lfloor{\alpha N}\right \rfloor$ most significant links, where $N$ is the total number of images in the collection (and $\alpha$ is a constant). 
	\item \textit{coverage} approach: we define graph coverage as the ratio of connected galleries pairs to the total number of possible gallery pairs. The latter is equal to $k(k-1)/2$, where $k$ is the number of galleries in the collection. We start by selecting the most significant link (i.e., the pair of photos with the highest visual similarity) and we compute the current graph coverage. We repeat this process, and when we reach certain values of graph coverage we stop collecting new links for the pairs of galleries, which have at least $\left \lfloor{\alpha N}\right \rfloor$ links. These values of graph coverage are arbitrarily set starting from $10\%$ (the first one) up to $100\%$ with a step of $10\%$. The intuition behind this procedure is that we select only the strong links when available, but we also explore weaker links for the pairs of galleries that do not exhibit strong links, in order to be able to synchronize as many galleries as possible.
\end{itemize} 
For all the experiments conducted in Section~\ref{sec:results}, we set $\alpha=0.1$ based on a set of preliminary evaluations that indicated good balance between processing speed and detection accuracy.

\subsection{Galleries Graph Construction}
\label{sec:graph}
\begin{figure}[t!]
	\centering
	%\captionsetup{justification=centering}
	\includegraphics[width=0.9\linewidth]{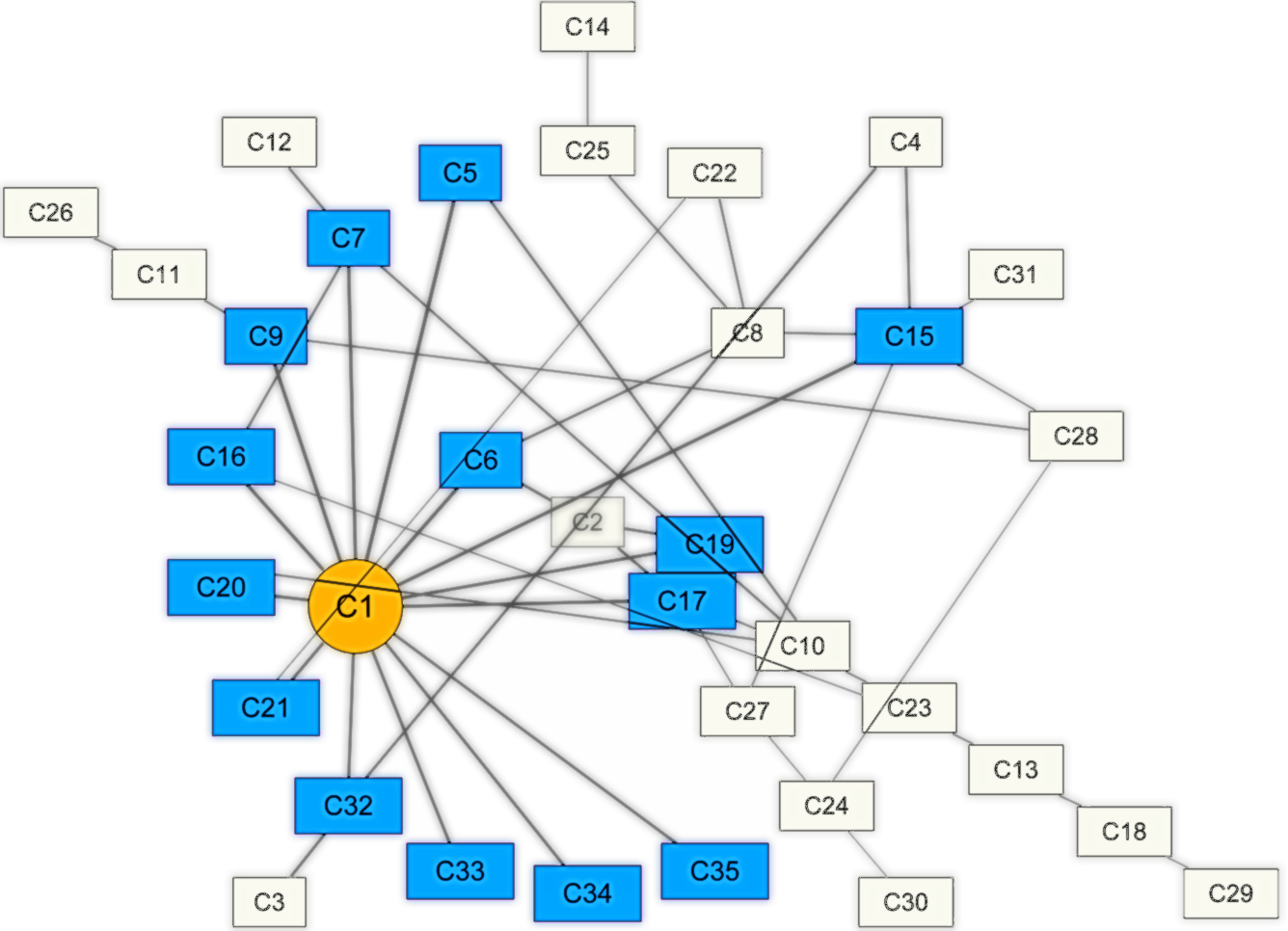}
	\caption{{Illustration of the galleries graph for the \textit{Vancouver} dataset used in our experiments.}}
	\label{fig:graphs}
\end{figure}
Having identified potential links for at least some gallery pairs, it is relatively straightforward to construct a weighted graph, whose nodes represent the galleries, and edges represent the links between galleries. The weight assigned to each edge is calculated as the median of the similarity values of the photos pairs that serve as links between the two galleries. {We follow this approach for calculating the edges' weights, in contrast to \cite{huevent15apo}, based on preliminary experiments, which showed that this can slightly improve the performance (compared to \cite{huevent15apo}) in three out of the four datasets employed in this work (see Section \ref{sec:framework} for details on the employed datasets).} Finally, using this graph, the temporal offsets of each gallery will be computed against one (random) gallery that is considered as the reference. Any gallery can be considered as the reference one, since we are aiming for relative synchronization; we neither assume nor need one gallery's timestamps to accurately match the true time.

It should be stressed here that both the \textit{exact} and \textit{coverage} approaches discussed in Section~\ref{sec:similarity} do not guarantee the discovery of links between the reference gallery and every other gallery, as shown in Fig.~\ref{fig:graphs}. In the figure, representing the galleries graph of one of the datasets used for evaluation, the first gallery is arbitrarily chosen as the reference gallery, and the corresponding node $C1$ is shown as an orange circle.
%
%For an indicative example see Fig.~\ref{fig:graphs}, where the galleries graph for one of the datasets used in this work is depicted. The first gallery of this dataset is arbitrarily chosen as the reference gallery in this example, and the corresponding node $C1$ is shown as an orange circle. 
Nodes corresponding to galleries that have direct links to the reference gallery, identified using the procedure described in Section~\ref{sec:similarity}, are depicted as blue rectangles. All other nodes of the graph are depicted as white rectangles. Attempting to directly compute the pairwise offsets between the reference gallery and each other gallery would not work, since only a portion of the galleries is directly connected with the reference gallery (regardless of which gallery was chosen as the reference, since the graph of Fig. \ref{fig:graphs} is not fully connected). Instead, we must find a way to properly traverse the graph in order to estimate an offset for each gallery.

\subsection{Temporal Offsets Estimation}
\label{sec:mrf}
Given the graph of galleries, the simplest approach to estimate the temporal offsets would be to use the method presented in \cite{huevent15apo} and traverse the minimum spanning tree (MST) of the galleries graph, synchronizing each pair of galleries using the median of the temporal offsets of all links connecting the two galleries. However, in order to improve the performance, we propose here a more elaborate approach based on probabilistic graphical models that substitutes the use of median, while still traversing the graph in the order specified by the MST (i.e., selecting pairs of galleries for offset estimation). 

Probabilistic graphical models allow to handle dependencies between random variables and to simultaneously take into account the uncertainty of inference. More specifically, we adopt undirected graphs, namely Markov Random Fields (MRFs), to model the similarities between photos as mutual dependence properties. %MRFs are used in a large variety of problems in image/video processing, including for example disparities computation in stereo vision applications~\cite{tappen2003comparison}, moving object detection in complex backgrounds~\cite{xu2012improved}, and image segmentation~\cite{hochbaum2001efficient}.

\begin{figure}[t!]
\centering{}
%\captionsetup{justification=centering}
\subfigure[]{\includegraphics[width=0.98\linewidth,trim=0cm 0.5cm 0cm 0.5cm]{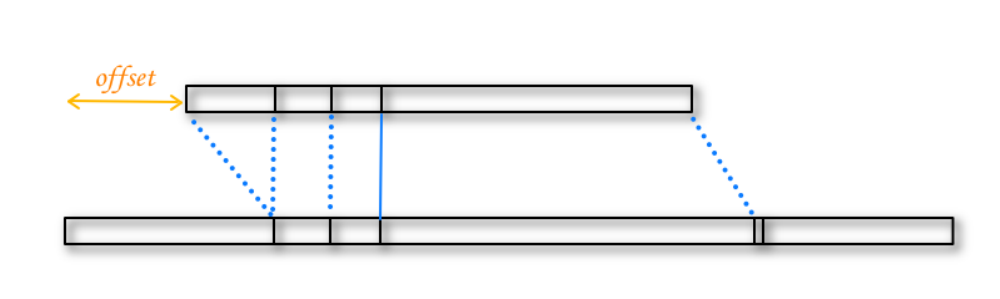}}
\subfigure[]{\includegraphics[width=0.98\linewidth,trim=0cm 0.5cm 0cm 0cm]{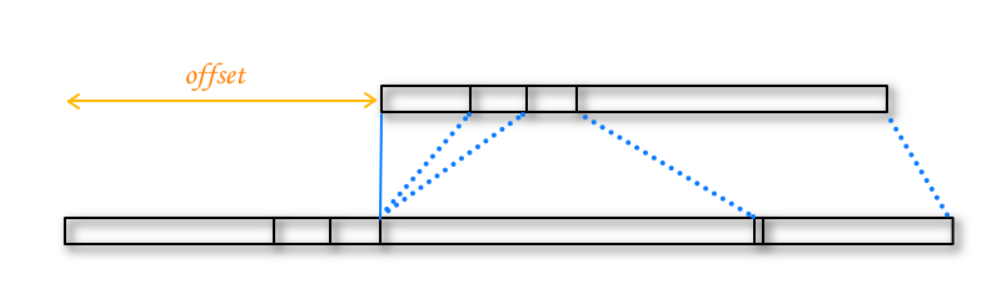}}
\caption{{Examples of correspondences between two galleries with different offsets. In each subfigure the lower stripe is the reference and the upper is the gallery to be synchronized; black vertical lines correspond to pictures, dotted and continuous blue lines are correspondences between photos.}}
\label{fig:sync}
\end{figure}

{From the minimum spanning tree, we compute the offset between galleries by finding the best correspondences of photos. In Fig.~\ref{fig:sync} black stripes and black vertical segments are used to symbolize galleries and photos, respectively. In each sub-figure, the lower stripe corresponds to the reference gallery, while the upper stripe is the gallery to be synchronized. The two examples depict two different situations with different offsets. Each of these offsets is defined according to a pair of visually similar images, which correspond to a potential link discovered at the previous stage (and visualized as continuous blue line). All other images are compared against their closest reference neighbors given that offset (shown as dotted blue lines). The set of all image comparisons is called the set of correspondences. It is evident from Fig.~\ref{fig:sync}, that the first offset produces a better set of correspondences than the second one, due to a better satisfaction of the temporal constraints imposed by the acquisition timestamps of images in both galleries. This definition of offsets, and thus correspondences, considerably differs from the model in~\cite{semMRF}. In fact, the possible offsets in~\cite{semMRF} are obtained by discretizing the temporal axis and by considering only those time instants that produce different sets of correspondences. Here, the offsets are defined based on the potential links obtained in the previous stage and the set of correspondences are defined according to each of these possible offsets. As mentioned previously, the advantage of this definition compared to the one proposed in~\cite{semMRF} is twofold: (i) a speedup of the inference stage due to a decreased set of possible solutions; (ii) the removal of the discretization parameter, which required in~\cite{semMRF} two different types of synchronization (a coarse and a fine synchronization).}

Notice that the number of images in each gallery in input to the graphical model is generally smaller compared to their original versions, since only visually similar images are now considered. In order not to complicate the notation, we keep denoting with $N_1$ and $N_2$ the size of the processed galleries (namely, the ones provided by the graph construction stage, see Section~\ref{sec:graph}).
Let us now assume that gallery $P^1$ is the reference and that $P^2$ has to be synchronized with respect to $P^1$. We start by defining the set of all possible temporal offsets as $\{\Delta T_m: m=1,\dots,Q\}$ ($Q$ is the cardinality of the set); these offsets are determined by the potential links (see Section \ref{sec:similarity}) of the two galleries, namely by the pairs of visually similar photos that have been found (the difference of timestamps between the photos in each pair is considered a possible temporal offset). Then, by defining the sequence of correspondences between the two galleries, and given the offset $\Delta T_m$, as $\mathbf{x}^{\Delta T_m}=(x^{\Delta T_m}_1,\dots,x^{\Delta T_m}_l,\dots,x^{\Delta T_m}_{N_2})$, where $x^{\Delta T_m}_l$ represents the photo in the reference gallery associated with photo $I_l^2$, the synchronization can be formulated as an optimization problem. In other words,

\begin{equation}
\Delta T^*=\arg \max_{\Delta T_m}f(\mathbf{x}^{\Delta T_m})
\label{eq:general}
\end{equation}
where $f:X\rightarrow\mathbb{R}$ is the function that assigns a similarity score to each sequence of correspondences and $X=\{\mathbf{x}^{\Delta T_m}:m=1,\dots,Q\}$ is the set of all possible candidate solutions. Therefore, the optimization consists of finding the best set of correspondences in order to maximize the similarity between galleries.

\begin{figure}[t!]
\centering
%\captionsetup{justification=centering}
\includegraphics[width=0.9\linewidth]{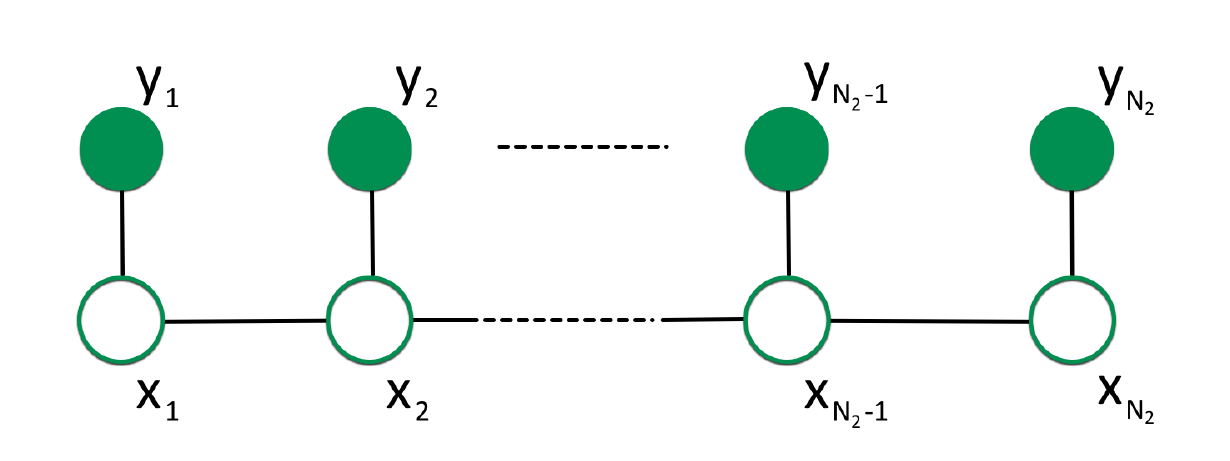}
\caption{Markov network with observed and hidden nodes.}
\label{fig:mkv}
\end{figure}

By introducing the sequence of observed nodes $\mathbf{y}=(y_1,\dots,y_i,\dots,y_{N_2})$, where $y_i$ refers to photo $I_i^2$ in $P^2$, and considering the sequence of latent variables $\mathbf{x}=(x_1,\dots,x_i,\dots,x_{N_2})$ with event space $X$, it is possible to define an undirected graphical model, as shown in Fig.~\ref{fig:mkv}. The edges of the graph serve two purposes: edges between nodes in $\mathbf{x}$ and $\mathbf{y}$ are used to compare images across the two galleries, while edges between nodes belonging to the sequence $\mathbf{x}$, are used to catch the temporal correlation between images within the same gallery.

Function $f$ in \eqref{eq:general} is defined as the conditional distribution of $\mathbf{x}$ given $\mathbf{y}$. The dependence properties defined by the graph in Fig.~\ref{fig:mkv} permit to factorize the distribution, namely considering a potential function for each maximal clique, where in our case a maximal clique consists of just a single edge in the graph:

\begin{equation}
f(\mathbf{x})\doteq p(\mathbf{x}|\mathbf{y})=\frac{1}{Z}\prod_i \psi(x_i,x_{i+1})\prod_k\phi(x_k,y_k).
\label{eq:potentials}
\end{equation}
In \eqref{eq:potentials}, $\psi(x_i,x_{i+1})$ and $\phi(x_k,y_k)$ are the potential functions associated to the edge connecting $x_i$ and $x_{i+1}$ and to the edge connecting $x_k$ and $y_k$, respectively, while $Z$ is the normalization constant.

\subsubsection{Definition of the potential functions}
\label{subsec:potential}
The potential $\phi(x_k,y_k)$ in~(\ref{eq:potentials}) quantifies the dissimilarity between pairs of photos across the two galleries according to localization information. Intuitively, photos acquired from the same location are more likely to refer to the same event. Conversely, photos acquired in completely different locations are more likely to be related to different events.

%{TO DO: remove description of features used by the older MRFsync method}
%As far as the visual similarity of photos is concerned, we adopt global descriptors related to the color and texture characteristics of a photo, in order to have a compact and global representation of the image content, that is as much generic as possible and is not restricted to a specific domain of events. For example, the method should be applicable to scenarios involving a huge number of people, such as concerts or international meetings, as well as to more personal situations such as weddings or holidays. To this end Color Structure Descriptors (CSD) are used to describe photos on a color basis. In comparison to standard color histograms, they also contain relevant information about the spatial structure of colors, leading to better performance in retrieval ranking systems~\cite{messing2001mpeg} compared to other descriptors defined in the MPEG-7 Standard. On the other hand, Local Binary Patterns (LBP) are used to assess the information about the texture of images~\cite{ojala1994performance}. The Euclidean distance is used in both cases to measure the visual dissimilarity; we denote as $D_C$ the distance between two images when using CSD and $D_L$ when using LBP.

When available, the GPS information stored in the photo headers is used to compare a pair of photos on the basis of the spatial distance between their acquisition locations. GPS coordinates, expressed in terms of latitude and longitude, can be transformed into spherical coordinates by assuming that the Earth is spheric. In fact, if $u_I^{G}$ and $v_I^{G}$ are the latitude and the longitude coordinates for photo $I$ and $u_I^{S}, v_I^{S}$ and $w_I^{S}$ are used to identify the correspondent spherical components, then it is possible to apply the following transformation:

\begin{equation}
\mathbf{z}_{I}=\begin{cases}
\begin{array}{c}
\begin{array}{l}
u_{I}^{S}=R\cos\left(\frac{2\pi u_{I}^{G}}{360}\right)\cos\left(\frac{2\pi v_{I}^{G}}{360}\right)\\
v_{I}^{S}=R\cos\left(\frac{2\pi u_{I}^{G}}{360}\right)\sin\left(\frac{2\pi v_{I}^{G}}{360}\right)\\
w_{I}^{S}=R\sin\left(\frac{2\pi u_{I}^{G}}{360}\right)\end{array}\\
\end{array}\end{cases}
\end{equation}
where $\textbf{z}_I=(u_{I}^{S},v_{I}^{S},w_{I}^{S})$ is introduced for notation compactness and $R$ is the average radius of the Earth. The orthodromic distance between the acquisition locations of two photos $I_1$ and $I_2$ can be therefore computed using the following relation:

\begin{equation}
D_{G}(I_1,I_2) = 2Rsin^{-1}\left(\frac{D(\mathbf{z}_{I_1},\mathbf{z}_{I_2})}{2R}\right)
\label{eq:gps}
\end{equation}
where $D(\mathbf{z}_{I_1},\mathbf{z}_{I_2})$ is the Euclidean distance between $\mathbf{z}_{I_1}$ and $\mathbf{z}_{I_2}$. In case images are not geo-tagged, the distance in \eqref{eq:gps} is set to $0$ and the localization information does not therefore contribute to the estimation of the offset.

The potential $\phi(x_k,y_k)$ in \eqref{eq:potentials} is defined as:

%In the previous paragraphs we have defined the metrics used to compare two images belonging to different galleries, according to the visual content and the GPS information. 
%Finally, we have to define the potential function associated to the link between an observed node and a hidden one. Considering that the most convenient way to define a potential is by using exponential functioIn fact, this method can be applied to ns, as indicated in \cite[eq. (8.41)]{bishop2006pattern}, we express the potential $\phi(x_i,y_i)$ by linearly combining the above three contributions, namely:

\begin{equation}
\begin{split}
\phi(x_{k},y_{k}) =& \exp\bigg\{-\gamma\frac{D_{G}(x_{k},y_{k})}{D_{G}^{max}(k)}\bigg\}
\end{split}
\label{eq:global}
\end{equation}
where $x_k$ and $y_k$ are the observed and hidden nodes of the MRF model, respectively, $D_{G}^{max}(k)=\max_{x_k}\left\{D_{G}(x_{k},y_{k})\right\}$ is the normalization term and $\gamma$ is a real positive parameter used to scale the importance of the orthodromic distance to offset estimation. %Intuitively, if GPS tags are available, then localization information provides great contribution in understanding if two photos belong to the same event and the associated weight has therefore a high value. Conversely, the weights associated to visual descriptors become dominant when localization information is missing.
%should be more meaningful in case  %with respect to visual aspects, thus $\gamma$ should be greater than $\alpha$ and $\beta$.
%Instead, we can not make any assumption on $\alpha$ and $\beta$, because we don't know if histograms approach is better than the SURF one. 
%These parameters could be set manually, but this option is not efficient, because the performances of the algorithm depend from their configuration and the manual setup requires to make an exhaustive search for finding the best solution. In order to avoid this, we have to learn the parameters in an automatic, but supervised way, which will be discussed later.

The potential $\psi(x_i,x_{i+1})$ in \eqref{eq:potentials} is instead used to account for the temporal dimension. In particular, it considers pairs of photos for both galleries and measures the quality of the alignment given a particular offset. The measure is based on the distance $D_{T}$ using a metric, namely the $l_1$-norm:
%The second potential that we need to define is the one that characterizes the connections among hidden nodes. It has to take into account the temporal component, preserving the temporal offset among photos of the same gallery. Both requirements are fulfilled by the following function:
%
\begin{equation}
D_{T}(x_{i},x_{i+1})=\bigg\|
\bigg[\begin{array}{c}
t_{y_i}\\
t_{y_{i+1}}
\end{array}\bigg]
{-}\bigg[\begin{array}{c}
t_{x_i}\\
t_{x_{i+1}}
\end{array}\bigg]
\bigg\|_1
\label{eq:time}
\end{equation}
where $t_{y_i}$ and $t_{y_{i+1}}$ denote the timestamps of the $i$-th and $i+1$-th image in $P^2$, while $t_{x_i}$ and $t_{x_{i+1}}$ are the timestamps of the corresponding images in $P^1$.
%Since $j = i + 1$, $t_{y_j}\geq t_{y_i}$, and $t_{x_j}\geq t_{x_i}$, we have that the two subtractions in the brackets of \eqref{eq:time} return positive quantities.
% Furthermore, if these values are equal, which means that the associations between photos is perfect from the temporal point of view, then we have that $D_{TIME}(I_{x_i} , I{x_j} ) = 0$.
If $t_{y_j}=t_{x_j}$ and $t_{y_i}=t_{x_i}$, then the image pairs are perfectly aligned and $D_{T}(x_i , x_{i+1}) = 0$. The potential $\psi(x_i,x_{i+1})$ can then be defined similarly to \eqref{eq:global}:
\begin{equation}
\psi(x_i,x_{i+1}) = exp\left\{-\delta\frac{D_{T}(x_i, x_{i+1})}{D_{T}^{max}(i)}\right\}
\label{eq:potentialtime}
\end{equation}
where $D_{T}^{max}(i)=\max_{x_i,x_{i+1}}\left\{D_{T}(x_i, x_{i+1})\right\}$ is the normali\-zation term and $\delta$ is a real positive parameter used to weight the temporal distance component.

It is worth mentioning that the potentials $\phi(x_{k},y_{k})$ and $\psi(x_i,x_{i+1})$ are defined so that the objective function in \eqref{eq:potentials} is maximized when the correspondences between photos across the two galleries have both strong location similarity and precise temporal alignment. The contribution to the offset estimation of these two terms can be weighted by setting differently the values of the potentials parameters, namely $\gamma, \delta$.

% Also in this case $j = i + 1$ is valid.

%With the definition of potential functions we have concluded the formulation of the graphical model for the photo galleries synchronization problem. 
%It is worth mentioning that the choice of the reference gallery within the whole set of galleries to be synchronized is non trivial. In fact, the reference has to be representative enough of the event, which means that not only its total duration should be the largest, but it should also represent the majority of the sub-events belonging to the same event. This will guarantee a sufficient overlapping between reference and actual gallery to allow an effective matching.

\subsubsection{Parameter estimation}
\label{subsec:learning}
Starting from a training set of galleries provided with the ground truth related to the synchronization offset, it is possible to estimate the parameters of the model described in the previous subsections. In fact, if $\bm{\theta}$ is the parameters vector, namely $\bm{\theta}\doteq[\gamma,\delta]^T$, and $\mathbf{x}^*$ is the sequence of visually similar photos between a pair of training galleries given the true offset (which can be obtained by using the stages preceding the MRF model), then the optimal parameters $\bm{\theta}^*$ are estimated as follows:

\begin{align}
\bm{\theta}^* &= arg\, max_{\bm{\theta}}\left\{ p(\mathbf{x}^*|\mathbf{y})\right\} \nonumber\\
 &= arg\, max_{\bm{\theta}}\left\{\cfrac{exp\left\{ -\sum_i\theta_ih_i(\mathbf{x}^*)\right\} }{\sum_{\mathbf{\tilde{\mathbf{x}}}}exp\left\{ -\sum_i\theta_ih_i(\tilde{\mathbf{x}})\right\}}\right\}
\label{eq:max}
\end{align}
where the second equality in \eqref{eq:max} is obtained by substituting \eqref{eq:global} and \eqref{eq:potentialtime} into \eqref{eq:potentials} and by exploiting the following relations:
\begin{align}
h_1(\mathbf{x}) &\doteq \sum_{k=1}^{N_2}\frac{D_{G}(x_{k},y_{k})}{D_{G}^{max}(k)}\nonumber\\
h_2(\mathbf{x}) &\doteq \sum_{i=1}^{N_2-1}\frac{D_{T}(x_{i},x_{i+1})}{D_{T}^{max}(i)}.
\label{eq:explfunction}
\end{align}
%In \eqref{eq:explfunction} each of the equations represents the contribution that a feature descriptor brings to the objective conditional distribution for a particular offset.

It is worth mentioning that, since the objective in \eqref{eq:max} is convex, the optimal parameters vector $\bm{\theta}^*$ can be estimated through standard approaches to convex optimization like the gradient descent algorithm. Another important aspect is that the model can be easily integrated with other types of features when available, including textual tags or data crawled from social networks. The wider the set of features, the more robust the model can be in estimating the offset.

\begin{figure}[t!]
	%\centering{}
	\includegraphics[width=0.95\linewidth]{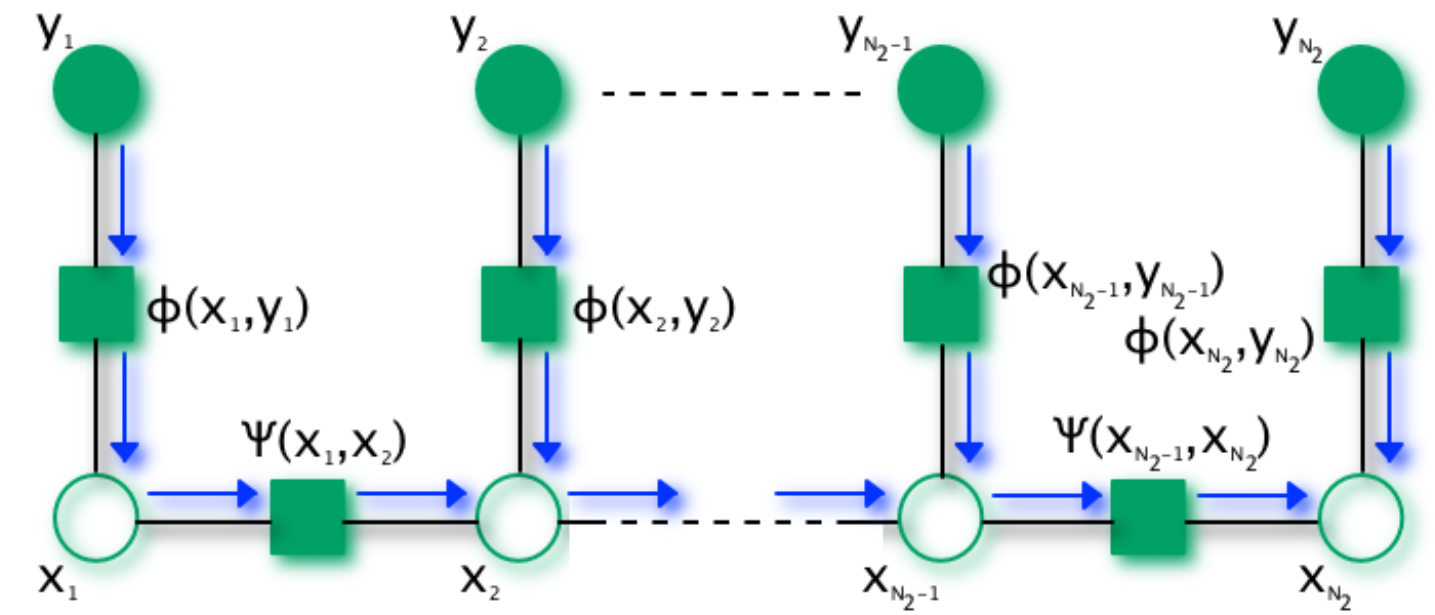}
	\caption{Equivalent factor graph.}
	\label{fig:factor}
\end{figure}

\subsubsection{Offset estimation}
\label{subsec:maxsum}

After the parameters of the potentials are estimated, the offset is found by computing the optimal sequence of states over the set $X$, namely finding:
\vspace{-3pt}
\begin{align}
\mathbf{x}_{max} &= arg\, max_{\mathbf{x}}\left\{ p(\mathbf{x}|\mathbf{y})\right\} \nonumber\\
&= arg\, max_{\mathbf{x}}\left\{\prod_i \psi(x_i,x_{i+1})\prod_k\phi(x_k,y_k)\right\}
\label{eq:estim}
\end{align}
where the second equality holds since the normalization constant $Z$ in \eqref{eq:potentials} does not affect the maximization. The expression above can be solved efficiently through exact inference thanks to the tree-structured shape of our graphical model. In particular, the max-sum algorithm is adopted: firstly, the Markov network has to be converted into an equivalent factor graph; secondly, messages are propagated from the leaves to the root, which can be chosen arbitrarily, and finally the maximization is performed. Fig. \ref{fig:factor} shows the exact flow of messages heading towards the root $x_{N_2}$, highlighted by blue arrows.

By following the approach in~\cite{bishop2006pattern} and by introducing $\mu_{a\rightarrow b}$ as the message propagating from a generic point $a$ of the graph to any other point $b$, it is possible to define the complete set of messages in the following way:

\begin{align}
\mu_{y_{k}\rightarrow\phi(x_{k},y_{k})}(y_{k})\doteq & 0\nonumber\\
\mu_{\phi(x_{k},y_{k})\rightarrow x_{k}}(x_{k})\doteq &\ln\left(\phi(x_{k},y_{k})\right)\nonumber\\
\mu_{x_{i}\rightarrow\psi(x_{i},x_{i+1})}(x_{i})\doteq &\mu_{\psi(x_{i-1},x_{i})\rightarrow x_{i}}(x_{i})+\\
& +\mu_{\phi(x_{i},y_{i})\rightarrow x_{i}}(x_{i})\nonumber\\
\mu_{\psi(x_{i-1},x_{i})\rightarrow x_{i}}(x_{i})\doteq &\ln\left(\psi(x_{i-1},x_{i})\right)+\nonumber\\
& +\mu_{x_{i-1}\rightarrow\psi(x_{i-1},x_{i})}(x_{i-1})\nonumber
\label{eq:messages}
\end{align}
where $k=1,\dots,N_2$ and $i=1,\dots,N_2-1$.

%According to the theory of the max-sum algorithm, the fourth equation in \eqref{eq:messages} should be defined as:
%
%\begin{equation}
%\begin{split}
%\mu_{\psi(x_{j-1},x_{j})\rightarrow x_{j}}(x_{j})=&\max_{x_{j-1}}\big\{ %\ln\left(\psi(x_{j-1},x_{j})\right)+ \\
%& +\mu_{x_{j-1}\rightarrow\psi(x_{j-1},x_{j})}(x_{j-1})\big\} 
%\end{split}
%\label{eq:messages2}
%\end{equation}
%
%This means that for each value of $x_{j}$ we need to store the best value of the previous node $x_{j-1}$, required during the back-tracking phase to obtain the best configuration of states. In our case, for each value of $x_{j}$  we have only one possible value for the previous node $x_{j-1}$, since the states are equivalent to possible time displacements between two galleries and have to be coherent among nodes. For this reason, we can omit the maximization operator, as used in \eqref{eq:messages2}, and consequently skip the back-tracking phase. 

Once all messages arrive to the root $x_{N_2}$, the best state, and consequently the best offset, is computed through the  maximization:

\begin{equation}
\begin{split}
x_{N_2}^{max}=&\arg\,\max_{x_{N_2}}\big\{ \ln\left(\phi(x_{N_2},y_{N_2})\right)+\\
& +\mu_{\psi(x_{N_2-1},x_{N_2})\rightarrow x_{N_2}}(x_{N_2})\big\}.
\end{split}
\label{eq:offset}
\end{equation}

\section{Experimental Results}
\label{sec:results}
In this section we present our experimental results. In \sec~\ref{sec:framework} we describe the datasets and the evaluation procedure we adopt. In \sec~\ref{sec:comparison_sims} we investigate the impact of different approaches used to identify the most similar images between different galleries and the impact of features extracted from different layers of the GoogLeNet DCNN used to assess visual similarity. Subsequently, in \sec~\ref{sec:comparison_others} we evaluate the effectiveness of the proposed method and compare it to state-of-the-art methods from the literature. Finally, in \sec~\ref{sec:comparison_ours} we discuss time complexity and storage requirements of the proposed method.

\begin{table}[t]
	\centering
	\caption{Datasets used for evaluation.}
	\label{tab:datasets}
	\begin{tabular}{|l|c|c|c|c|}
		\hline
		& \textit{London} & \textit{Vancouver} & \textit{NAMM15} & \textit{TDF14} \\ \hline \hline
		Number of photos    & 1351       & 2124        & 420       & 2471    \\ \hline
		Number of galleries & 35         & 37          & 19        & 33      \\ \hline
	\end{tabular}
\end{table}

\setlength\extrarowheight{2pt}
\begin{table*}[t!]
	\centering
	\caption{Results of using the different DCNN-based features (discussed in Section~\ref{sec:similarity}) and the different links identification approaches (discussed in Section~\ref{sec:graph}).}
	\begin{tabular}{|P{1.6cm}|P{1.3cm}|P{0.7cm}P{0.7cm}|P{0.7cm}P{0.7cm}|P{0.7cm}P{0.7cm}|P{0.7cm}P{0.7cm}|P{1.3cm}P{1.3cm}P{1.3cm}|}\hline
		GoogLeNet Layer Name & 
        Links identif. approach &
        \multicolumn{2}{c|}{\textit{Vancouver}} & 
        \multicolumn{2}{c|}{\textit{London}} & 
        \multicolumn{2}{c|}{\textit{NAMM15}} & 
        \multicolumn{2}{c|}{\textit{TDF14}} & 
        \multicolumn{3}{P{4.9cm}|}{Average and standard deviation across all datasets} \\
		& &P(\%)&A(\%)&P(\%)&A(\%)&P(\%)&A(\%)&P(\%)&A(\%)&P(\%)&A(\%)&H(\%)\\ \hline \hline

        \multirow{2}{2.8cm}{loss3/\\classifier}
        & \textit{coverage} &26.5&94.0&36.1&81.1&72.2&67.5&25.0&75.0&40.0$\pm$22.1&79.4$\pm$11.2&49.7$\pm$14.4\\ \cline{2-13}
		& \textit{exact} &5.9&56.9&44.4&64.3&77.8&76.9&43.8&45.2&43.0$\pm$29.4&60.8$\pm$13.3&46.3$\pm$27.6\\ \hline

       	\multirow{2}{2.8cm}{conv2/\\norm2}
        & \textit{coverage} &91.2&82.8&25.0&77.1&\textbf{83.3}&92.8&62.5&84.1&65.5$\pm$29.6&84.2$\pm$6.5&71.0$\pm$23.4\\ \cline{2-13}
		& \textit{exact} &76.5&63.4&\textbf{77.8}&78.8&\textbf{83.3}&91.6&\textbf{65.6}&72.9&75.8$\pm$7.4&76.7$\pm$11.8&76.0$\pm$8.7\\ \hline

		\multirow{2}{2.8cm}{inception3a} 
        & \textit{coverage} &94.1&90.5&41.7&79.5&\textbf{83.3}&90.8&\textbf{65.6}&81.5&71.2$\pm$22.9&85.6$\pm$5.9&76.6$\pm$16.8\\ \cline{2-13}
		& \textit{exact} &\textbf{97.1}&83.7&75.0&84.3&\textbf{83.3}&93.4&\textbf{65.6}&73.7&\textbf{80.3}$\pm$13.3&83.8$\pm$8.0&\textbf{81.7}$\pm$9.4\\ \hline

		\multirow{2}{2.8cm}{inception4a} 
        & \textit{coverage} &94.1&86.7&30.6&79.3&\textbf{83.3}&91.0&\textbf{65.6}&75.4&68.4$\pm$27.8&83.1$\pm$7.0&72.9$\pm$21.1\\ \cline{2-13}
		& \textit{exact}    &85.3&82.1&38.9&85.2&\textbf{83.3}&85.3&\textbf{65.6}&70.7&68.3$\pm$21.5&80.8$\pm$6.9&72.4$\pm$14.7\\ \hline

		\multirow{2}{2.8cm}{inception5a} 
        & \textit{coverage} &\textbf{97.1}&\textbf{96.0}&36.1&77.1&77.8&90.7&50.0&78.1&65.2$\pm$27.4&85.5$\pm$9.3&72.6$\pm$21.5\\ \cline{2-13}
		& \textit{exact} &73.5&76.7&41.7&84.3&\textbf{83.3}&92.8&56.3&\textbf{92.8}&63.7$\pm$18.5&86.6$\pm$7.8&72.2$\pm$13.3\\ \hline

		\multirow{2}{2.8cm}{Fusion\\approach} 
        & \textit{coverage} &\textbf{97.1}&88.4&25.0&\textbf{95.0}&\textbf{83.3}&\textbf{96.9}&\textbf{65.6}&75.7&67.8$\pm$31.3&\textbf{89.0}$\pm$9.6&73.0$\pm$24.4\\ \cline{2-13}
		& \textit{exact} &91.2&79.2&75.0&84.6&77.8&87.5&\textbf{65.6}&79.2&77.4$\pm$10.6&82.6$\pm$4.1&79.6$\pm$5.6\\ \hline
	\end{tabular}
	\label{tab:results_sims}
\end{table*}

% \begin{table}[t!]
% 	\centering
% 	\caption{}
% 	\begin{tabular}
%     {|P{1.0cm}|P{0.5cm}P{0.5cm}|P{0.5cm}P{0.5cm}|P{0.5cm}P{0.5cm}|P{0.5cm}P{0.5cm}|}\hline
% \cite{huevent15apo}   	& 97.1 & 86.0 & 63.9 & 75.0 & 50.0 & 71.5 & 21.9 & 90.8\\ \hline
% \cite{huevent15apo}* 	& 97.1 & 83.4 & 63.9 & 78.5 & 73.6 & 85.8 & 59.3 & 44.4\\ \hline
% 	\end{tabular}
% 	\label{tab:results_counts_vs_sims}
% \end{table}

\subsection{Datasets and Evaluation Framework}
\label{sec:framework}
For the experimental validation of the proposed methods we employ the two datasets of the MediaEval 2014 SEM task \cite{conci:sem}, consisting of photos from various users taken during two Olympic Games events, and two datasets of the MedialEval 2015 SEM task \cite{sem15_overview}, consisting of photos from an exhibition and a cycling event, for a total of four different datasets\footnote{The MediaEval Development Sets and Test Sets include material subject to Creative Commons license and are freely available for download at http://mmlab.disi.unitn.it/MediaEvalSEM2014 and http://mmlab.disi.unitn.it/MediaEvalSEM2015.}. All photos for each of these datasets are organized in galleries (each gallery captured using a single device) and come with timestamps, which are consistent within each gallery but may have considerable temporal offsets across different galleries. Furthermore, some galleries include geo-location information, while others do not. We strictly followed the experimental setup and evaluation procedure of the MediaEval SEM tasks, making our results directly comparable to the results reported by the tasks participants.

The first test dataset, \textit{Vancouver}, is about the Winter Olympics Games held in 2010, consisting of 1351 photos arranged in 35 galleries, while the second, \textit{London}, is about the Olympic Games held in 2012 and consists of 2124 photos arranged in 37 galleries. The third dataset concerns the famous exhibition held every year in California for music merchants (\textit{NAMM15}), consisting of 420 images and 32 videos, split into 19 galleries with each user gallery containing a variable number of media. In the context of this work we deal only with the images included in the dataset, for simplicity (but our algorithm can be easily extended to take into account videos files as well). Finally, the fourth dataset is related to the Tour de France event held in 2014 (\textit{TDF14}). The dataset is split into 33 galleries and covers the entire competition. As part of the adopted experimental setup, in all datasets the first gallery is considered as the reference one. Table \ref{tab:datasets} summarizes the four datasets that we use.

As far as the evaluation metrics are concerned, we have adopted the \textit{precision} and \textit{accuracy} to assess the quality in synchronization, as defined in \cite{conci:sem}, and briefly reported hereafter for convenience.

\textbf{1) Precision} ($P$) corresponds to the ratio between the number of synchronized galleries ($M_{syn}$), and the total number of galleries ($M - 1$, excluding the reference gallery) in the dataset. A gallery is considered to be synchronized if the difference between the estimated timestamps of its photos and the corresponding ground truth is lower than a maximum accepted temporal offset ($maxError$). The value of $maxError$ is set to $1800$ seconds as per the guidelines in \cite{conci:sem}. The Precision measure is defined as:
\begin{equation}
P = \frac{M_{syn}}{M-1}.
\end{equation}

\textbf{2) Accuracy} ($A$) is the average temporal offset calculated over the synchronized galleries, normalized with respect to the maximum accepted temporal offset ($maxError$). The synchronization error for  gallery $P^i$ with respect to the reference gallery $r$ is defined as $\Delta E_{ir} = | \Delta T_{ir} - \Delta T_{ir}^* |$, where $\Delta T_{ir}$ and $\Delta T_{ir}^*$ are the offset between gallery $i$ and gallery $r$ obtained from the method and the ground truth, respectively. Thus, the Accuracy measure is defined as: 
\begin{equation}
A = 1 - \frac{\sum_{i=1}^{M_{syn}}\Delta E_{ir}}{M_{syn} \cdot \operatorname{maxError}}.
\end{equation}

\textbf{3) Harmonic mean} ($H$). We combine the aforementioned measures according to: 
\begin{equation}
H = (2 \! \cdot \! P \! \cdot \! A)/(P \! + \! A).
\end{equation}

{In the subsequent experiments, we also report the average and standard deviation of each of the aforementioned evaluation metrics, across the four datasets used.}

\subsection{Impact of Different Design Choices}
\label{sec:comparison_sims}
In order to understand the impact of different approaches for constructing the galleries graph, as discussed in \sec~\ref{sec:similarity}, we conducted experiments on:
\begin{enumerate}
	\item The different DCNN layers to extract features from ("loss3/classifier", "conv2/norm2", "inception3a/output", "inception4a/output" and "inception5a/output"), as well as their combination by means of late fusion,
	\item The two alternative approaches for identifying the most similar images per gallery pair (\textit{exact}, \textit{coverage}).
\end{enumerate}

The results are shown in Table~\ref{tab:results_sims}, where we report the synchronization precision and accuracy for each dataset, as well as the averages across all datasets for all three metrics defined in Section~\ref{sec:framework} (precision, accuracy, harmonic mean). The best results for each column of this table are shown in bold. According to these results, the "loss3/classifier" layer's features exhibit the worst performance, since this kind of features assesses the purely semantic (rather than visual) similarity of images. In contrast, the best overall results are achieved when using the "inception3a/output" layer's features and the \textit{exact} approach: the harmonic mean reaches $81.7\%$, the precision also receives its highest value (at $80.3\%$), and accuracy achieves a score of $83.8\%$. Considering these results, we can conclude that, in our implementation, the "inception3a/output" layer and the \textit{exact} approach, out of those discussed in Section~\ref{sec:similarity}, represent the best choices for evaluating photo similarity.

\subsection{Comparison with the State-of-the-Art}
\label{sec:comparison_others}

\begin{table*}[ht]
\centering
\caption{Comparison between the proposed method and the methods of \cite{broilo:icip,zaharieva2014multimodal,semJRS,apostolidis2014certh,semMRF,huevent15apo,sem15jrs,sem15certh}.}
\begin{tabular}{|P{3.0cm}|P{0.7cm}P{0.7cm}|P{0.7cm}P{0.7cm}|P{0.7cm}P{0.7cm}|P{0.7cm}P{0.7cm}|P{1.3cm}P{1.3cm}P{1.3cm}|}
\hline
&\multicolumn{2}{c|}{\textit{Vancouver}}
&\multicolumn{2}{c|}{\textit{London}}
&\multicolumn{2}{c|}{\textit{NAMM15}}
&\multicolumn{2}{c|}{\textit{TDF14}}
&\multicolumn{3}{P{4.9cm}|}{Average and standard deviation across all datasets} \\
Method & P(\%)&A(\%)&P(\%)&A(\%)&P(\%)&A(\%)&P(\%)&A(\%)&P(\%)&A(\%)&H(\%)\\ \hline \hline
\cite{broilo:icip} 				& 61.8 & 81.6 & 33.3 & 88.6	& 50.0* & 73.5* & 25.0 & 80.6 & 43.9$\pm$17.6& 81.1$\pm$6.2& 55.5$\pm$15.9\\ \hline
\cite{zaharieva2014multimodal}  & 94.1 & 79.2 & 47.2 & 87.5	& -    & -    & -    & -    & 70.7$\pm$33.2 & 83.4$\pm$ 5.9 & 73.6$\pm$17.2 \\ \hline
\cite{semJRS}    				&  5.0 & 65.0 & 15.0 & \textbf{92.0} & -    & - 	  & -    & -    & 10.0$\pm$ 7.1 & 78.5$\pm$19.1 & 17.6$\pm$11.6 \\ \hline
\cite{apostolidis2014certh}  	& 91.2 & 72.8 & 61.1 & 71.3	& 77.9* & 90.4* & 9.4 & 79.3 & 61.4$\pm$37.0 & 78.5$\pm$ 8.7 & 62.7$\pm$31.8 \\ \hline
\cite{sem15jrs}					& -    & -    & -    & -    & 40.0 & 78.0 & 5.0 & \textbf{92.0} & 22.5$\pm$24.5 & 85.0$\pm$ 9.9 & 31.2$\pm$30.6 \\ \hline
\cite{sem15certh}				& 94.1 & 76.0 & 61.1 & 65.6 & 83.3 & 90.8 & 12.5 & 84.5 & 62.8$\pm$36.2 & 79.2$\pm$10.9 & 64.0$\pm$30.1  \\ \hline
\cite{semMRF}    				& 35.0 & 86.0 & 25.0 & 89.0	& 44.5* & 82.4* & 15.6 & 83.2 & 30.0$\pm$12.5 & \textbf{85.2}$\pm$ 3.0 & 43.2$\pm$13.7  \\ \hline
\cite{huevent15apo}    			& \textbf{97.1} & 86.0 & 63.9 & 75.0	& 50.0* & 71.5* & 21.9 & 90.8 & 59.6$\pm$30.9 & 80.8$\pm$ 9.1 & 64.5$\pm$23.0  \\ \hline

{\cite{semMRF} modified (inception3a)}
								& 5.9 & \textbf{93.4} & 5.6 & 73.5 & 47.1 & 85.0 & 9.4 & 73.2 & 17.0$\pm$20.1 & 81.3$\pm$9.8 & 24.7$\pm$24.1\\ \hline
\cite{huevent15apo} modified (inception3a,\textit{exact})
								& 85.3 & 56.3 & 47.2 & 74.6 & \textbf{88.9}* & 88.7* & \textbf{65.6} & 84.6 & 73.1$\pm$20.9 & 76.1$\pm$14.4 & 72.7$\pm$14.1 \\ \hline
{\cite{huevent15apo} and \cite{semMRF}} & 8.8 & 54.7 & 19.4 & 67.4 & 58.8 & 66.8 & 25.0 & 80.6 & 28.0$\pm$21.6 & 67.4$\pm$10.6 & 36.5$\pm$19.8\\ \hline
Proposed (inception3a,\textit{exact},MRF)
			&\textbf{97.1}&83.7&\textbf{75.0}&84.3&83.3*&\textbf{93.4}*&\textbf{65.6}&73.7&\textbf{80.3}$\pm$13.3&83.8$\pm$8.0&\textbf{81.7}$\pm$9.4\\ \hline
\multicolumn{12}{P{17cm}}{* Note: since the 19-th gallery in the \textit{NAMM} dataset contains only videos, and for simplicity in this work we deal only with images included in the dataset, we regard this gallery as non-correctly synchronized in all the experiments we conducted.}
% 	MRF (Conv3a + Exact) 		& \textbf{97.1} & \textbf{83.7} & 75.0 & \textbf{84.3} & \deleted{88.2} & \textbf{93.4} & \textbf{65.6} & 73.7 & \textbf{81.5}$\pm$13.9 & \textbf{83.8}$\pm$\textbf{8.1} \\ 		
%     Median (Conv3a + Exact)  		    & 85.3 & \deleted{56.3} & \deleted{47.2} & 74.6 & \textbf{94.1} & \deleted{88.7} & \textbf{65.6} & \textbf{84.6} & \deleted{73.1}$\pm$\deleted{20.9} & \deleted{76.1}$\pm$\deleted{14.4} \\ 
%     MRF (Conv2a + Exact) 		& \deleted{76.5} & 63.4 & \textbf{77.8} & 78.8 & \deleted{88.2} & 91.6 & \textbf{65.6} & \deleted{72.9} & 77.0$\pm$\textbf{9.2} & 76.7$\pm$11.8 \\ 	
%     Median (Conv2a + Exact)  		    & \textbf{97.1} & 75.0 & 61.1 & \deleted{63.8} & \textbf{94.1} & 92.1 & \textbf{65.6} & 82.2 & 79.5$\pm$18.7 & 78.3$\pm$12.0 \\     		
% \hline
	\end{tabular}
	\label{tab:results_others}
\end{table*}

We compare the proposed method against the most relevant approaches of the literature \cite{broilo:icip,zaharieva2014multimodal,semJRS,apostolidis2014certh,sem15jrs,sem15certh} and also against the preliminary works \cite{semMRF,huevent15apo} that the proposed method is based on. To demonstrate the superiority of the MRF-based technique for the calculation of the galleries' offsets, we also include in the comparison: i) a modified version of our earlier method of \cite{huevent15apo} using the DCNN-based features; ii) a similarly modified version of~\cite{semMRF}; iii) a straightforward combination (cascade) of~\cite{huevent15apo} and~\cite{semMRF}, in which the MST-based procedure of~\cite{huevent15apo} is replaced by the original MRF-based method of~\cite{semMRF} in order to estimate the temporal offsets.
For the literature works that reported results of different variants of the algorithm, we include in Table \ref{tab:results_others} only the best results. 
As already mentioned, we strictly followed the experimental setup of the MediaEval SEM task and used the provided datasets and ground truth annotations to compare directly with the works of \cite{zaharieva2014multimodal,apostolidis2014certh,semJRS,sem15jrs,sem15certh}. To compare with \cite{broilo:icip}, we re-implemented the method and tested it under the same experimental setup. For methods \cite{zaharieva2014multimodal,semJRS,sem15jrs} we report the published results; thus, for \cite{zaharieva2014multimodal,semJRS} we report the results for the \textit{Vancouver} and \textit{London} datasets, while for \cite{sem15jrs}, we report the results for the \textit{NAMM15} and \textit{TDF14} datasets. The best results for each column in this table are shown in bold.

\begin{figure*}[t!]
	\centering{}
	% \captionsetup{justification=centering}
	\subfigure[]{\includegraphics[width=0.53\textwidth]{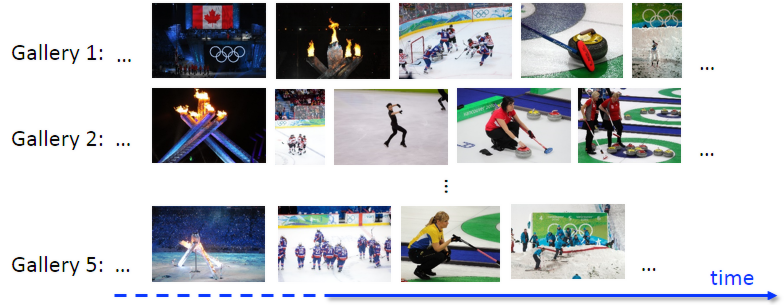}}\\
	\subfigure[]{\includegraphics[width=0.53\textwidth]{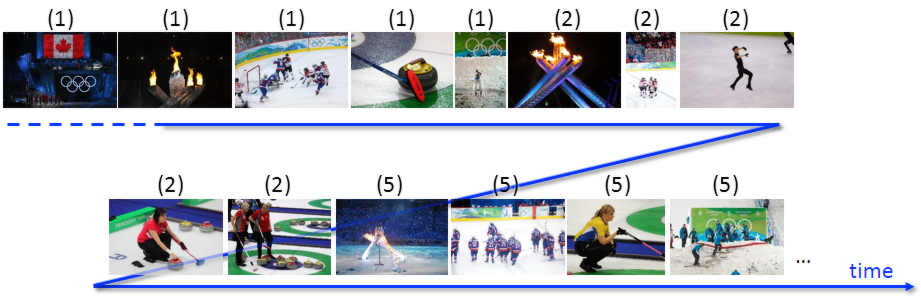}}\\
	\subfigure[]{\includegraphics[width=0.53\textwidth]{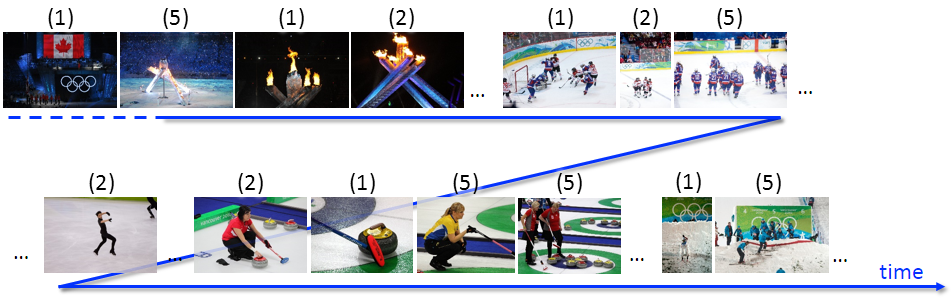}}       
	\caption{Qualitative example of synchronization for a small subset of photos from the \textit{Vancouver} dataset. (a) Original photo galleries. (b) Temporal order of the photos according to the original (noisy) timestamps. (c) Temporal order of the same photos using the timestamps estimated by the proposed method; these corrected timestamps make straightforward the meaningful ordering of the photos (opening ceremony, first ice hockey game, etc.) based on time information. In (b) and (c), the number above each photo indicates its gallery membership.}
	\label{fig:qual_example}
\end{figure*}

%\textcolor{red}{
	According to these results, the proposed method achieves the best precision in almost all datasets and outperforms on average all other approaches in the literature (\cite{broilo:icip,zaharieva2014multimodal,apostolidis2014certh,semJRS,sem15jrs,sem15certh}). Regarding accuracy, \cite{semJRS} scored the best accuracy in the \textit{London} dataset but managed to synchronize  a very small portion of the galleries. The same holds for the method of \cite{sem15jrs} and the \textit{TDF14} dataset. Jointly considering precision and accuracy, by calculating the harmonic mean measure across all datasets (last column of Table~\ref{tab:results_others}), reveals that the proposed method performs the best in comparison to all literature works, by a large margin.
%}

%\textcolor{red}{
	Concerning \cite{semMRF}, the proposed method significantly outperforms it, primarily due to \cite{semMRF} managing to synchronize only a small number of galleries (precision ranging from $15.6\%$ to $44.5\%$ for the different datasets). This continues to be the case when \cite{semMRF} is modified so as to use the DCNN-based	features discussed in Section~\ref{sec:similarity} instead of the original	ones; following this 	modification, the accuracy improves in two out of the four datasets but precision	drops considerably for three datasets. The reason for this is that in \cite{semMRF} no pre-filtering is foreseen and the features are directly used in the graphical model together with location and time. Instead, the adoption of the new features leads to considerably improved results when combined with the graph construction and navigation approach proposed in the present work.  Similarly \cite{huevent15apo} also performs worse than the proposed method (with average $H=64.5\%$ versus $H=81.7\%$ for the proposed). The modified version of \cite{huevent15apo}, that also uses the powerful DCNN-based features instead of the original ones, performs slightly worse on two datasets and significantly better on the third and fourth ones (NAMM15 and TDF14), compared to the original [38]. Overall, considering the average harmonic mean across all datasets, the performance of the modified \cite{huevent15apo} is better ($H=64.5\%$ versus $H=72.7\%$). The fluctuations in performance across datasets are attributed to the reliance of \cite{huevent15apo} and modified \cite{huevent15apo} methods to a multitude of thresholds, which cannot be tuned to a single set of values that is optimal for all datasets. These two comparisons highlight the fact that simply introducing the DCNN features in a synchronization method is not enough to obtain satisfactory results. The excellent performance of the proposed method is due in part to these features (with the harmonic mean value increasing from $64.5\%$ for the original method of \cite{huevent15apo}, to $72.7\%$ for the variant of it using our new features), and to an even greater extent to the MRF-based temporal offset estimation (proposed in Section~\ref{sec:mrf}) which exploits the information extracted from previous stages (with the average harmonic mean further increasing from $72.7\%$ to $81.7\%$).
%}

%\textcolor{red}{
	The straightforward combination of our preliminary methods \cite{huevent15apo} and \cite{semMRF} performs poorly, especially in terms of precision, which means that only few galleries are correctly synchronized. This is mainly due to the fact that \cite{semMRF} relies on the assumption that all galleries must have a considerable overlap with the reference gallery (i.e., there should be several links from the reference to all other galleries). It is clear that when straightforwardly combining \cite{huevent15apo} and \cite{semMRF}, there is no guarantee that a reasonably high number of links (or even at least one link) will be obtained for each pair of galleries. Due to this lack of links, the offset error propagated along the graph is higher than using our proposed approach, and the number of galleries correctly synchronized is very low.
%}

A qualitative example of the temporal synchronization achieved by the proposed method is shown, for a small subset of photos from the \textit{Vancouver} dataset, in Fig.~\ref{fig:qual_example}.

\subsection{Computational Complexity Concerns}
\label{sec:comparison_ours}
In this subsection, we derive the time and storage complexi\-ty required by our algorithm. To simplify the analysis without loosing generality, we assume that all galleries have the same number of photos, namely $\bar{N}$.\footnote{$\bar{N}$ can be regarded as the average number of photos per gallery, if one wants to relax this assumption.} Therefore, the total number of photos is $n=k\bar{N}$.

For simplicity, in the subsequent tables and discussion we use the following notation for the different stages of the proposed method's pipeline:
\begin{enumerate}
	\item \textbf{FE}: Feature extraction (Section~\ref{sec:similarity})
	\item \textbf{SIM}: Construction of the visual similarity matrix of all images (Section~\ref{sec:similarity})
	\item \textbf{GL}: Identification of the most similar photos between different galleries (Section~\ref{sec:similarity})
	\item \textbf{GC}: Construction of the galleries graph (Section~\ref{sec:graph})
	\item \textbf{MST}: Finding of the MST in the galleries graph (Section~\ref{sec:graph}) 
	\item \textbf{MRF}: Calculation of the temporal offsets using MRF (Section~\ref{sec:mrf})
\end{enumerate}
In Table~\ref{tab:times} we report the time needed (in seconds) for each of the aforementioned stages, on a Windows 7 64-bit machine with Intel i7 processor and 16GB of RAM.

\begin{table}[!ht]
	\centering
	\caption{Empirical time measurements (in seconds) of all stages of the proposed method.}
	\label{tab:times}
	\begin{tabular}{|p{1.3cm}|P{1.1cm}|P{1.1cm}|P{1.1cm}|P{1.1cm}|} \hline
    &\textit{Vancouver}&\textit{London}&\textit{NAMM15}&\textit{TDF14}\\ \hline \hline
    \textbf{FE}		& $1330.8$ & $1890.2$ & $387.3$ & $2162.5$ \\ \hline
    \textbf{SIM}    & $2.4$ & $5.7$ & $0.2$ & $7.9$ \\ \hline
    \textbf{GL}		& $0.2$ & $0.2$ & $0.1$ & $0.4$ \\ \hline	
    \textbf{GC}  	& $<0.1$ & $<0.1$ & $<0.1$ & $<0.1$ \\ \hline	
    \textbf{MST}	& $<0.1$ & $<0.1$ & $<0.1$ & $<0.1$ \\ \hline			
    %\textbf{MRF}	& 11.0 & 15.9 & 4.6 & 10.8 \\ \hline
    %\textbf{MRF}	& 7.4 & 11.4 & 3.0 & 7.6 \\ \hline
     \textbf{MRF}	& 6.1 & 9.3 & 2.5 & 6.2 \\ \hline
	\end{tabular}
\end{table}

\begin{table}[!ht]
	\centering
	\caption{Time and storage complexity of all stages of the proposed method with respect to the number of galleries ($k$), the number of photos in each gallery ($\bar{N}$), and the total number of photos in the collection ($n$).}
	\label{tab:complexities}
	\begin{tabular}{|p{1.3cm}|P{2.5cm}|P{2.5cm}|} \hline
					& Time Complexity & Storage Complexity\\ \hline \hline
	\textbf{FE}		&  $O(n)$ 		&  $O(n)$ 	\\ \hline
	\textbf{SIM}    &  $O(n^2)$ 	&  $O(\alpha n^2)$ \\ \hline
	\textbf{GL}		&  $O(k^2)$ 	&  $O(k^2)$ \\ \hline	
	\textbf{GC}  	&  $O(k^2)$     &  $O(k^2)$ \\ \hline	
	\textbf{MST}	&  $O(\frac{k^2}{\log{k}})$ & $O(k)$ \\ \hline		
	\textbf{MRF}	&  $O(k\bar{N}^3)$ 	& $O(k\bar{N}^2)$ \\ \hline
	\end{tabular}
\end{table}

Concerning time and space complexity (Table~\ref{tab:complexities}), it is easy to derive the worst case for the \textbf{FE}, \textbf{SIM}, \textbf{GL} and \textbf{GC} stages. Regarding the computation of the MST of the galleries graph (\textbf{MST} stage), we consider that: if $E$ is the number of edges and $V$ is the number of vertices in a graph, the Kruskal algorithm, which we employ is known to run in $O(E\log{V})$ time. Since, in our case $V=k$ and $E\leq k^2$ the time complexity is $O(\frac{k^2}{\log{k}})$. The storage complexity of this stage is $O(k)$, since in the worst case one needs to hold all vertices (galleries) in the queue. In the \textbf{MRF} stage, inference is performed through the max-sum algorithm described in \sec~\ref{subsec:maxsum}. Also in this case it is worth noting that the number of possible states/offsets for $\mathbf{x}$ is at most $\eta=\bar{N}^2$, but this number tends to be smaller when good matches across galleries are found. Similarly to \cite{bishop2006pattern}, one can easily derive the time and storage complexities of the max-sum algorithm for our graphical model, namely $O(\eta(\bar{N}-1))$ and $O(\eta)$, respectively. 

We notice from the data reported in Table~\ref{tab:complexities} that the \textbf{SIM} stage exhibits the highest complexity if the number of users is large. {Especially if the number of photos increases significantly the pairwise photo comparisons stage could consume considerable computational resource. However, the complexity of this stage can be reduced by using known data indexing schemes, such as Locality Sensitive Hashing and KD-trees, to avoid computing the similarities for all $n^2$ possible pairs of photos.}  The \textbf{MRF} stage is more computationally expensive in case of very large photo collections. 
%We notice from the data reported in Table~\ref{tab:complexities} that the \textbf{SIM} stage exhibits the highest complexity, while the rest of the stages present sub-quadratic complexity. \deleted{NOT TRUE, THERE IS A CUBIC TERM FOR THE MRF. MAYBE, WE SHOULD DISTINGUISH BETWEEN COMPLEXITY DEPENDING ON NUMBER OF USERS $k$ AND COMPLEXITY DEPENDING ON AVERAGE NUMBER OF PHOTOS $\bar{N}$, HIGHLIGHTING THE FACT THAT FOR SCENARIOS WITH LARGE NUMBER OF USERS THE BOTTLENECK OF OUR APPROACH IS IN THE SIM, GL AND GC STAGES, WHILE IN THE CASE OF LARGE PHOTO COLLECTIONS THE BOTTLENECK CORRESPONDS TO THE MRF STAGE.} 
Nevertheless, it is clear from our empirical time measurements (Table~\ref{tab:times}), that even stages with relatively high complexity have very short running times in practice.

\section{Conclusions}
\label{sec:conclusions}
In this paper we presented a method addressing the pro\-blem of synchronizing multiple photo galleries. %For this, we combined the (noisy) photo metadata related to time and geo-location together with the photos visual content. 
The proposed approach exploits Deep Convolutional Neural Network features to measure photo similarity; subsequently, similar photos are used to define a graph of galleries, and finally each pair of galleries in the graph is synchronized using a probabilistic graphical model.
%The method we proposed exhibits multiple advantages (scalability, synchronization precision and accuracy) compared to the state of the art. \deleted{WE NEED TO BE VERY CAREFUL TO TALK ABOUT SCALABILITY, SINCE THE DATASETS WE ARE USING ARE NOT LARGE. THIS ISSUE WAS ALSO RAISED BY A REVIEWER.}
%, making it suitable for use in such synchronization problems. 
Extensive experiments on four benchmark datasets documented the merit of the proposed method, which achieves very good synchronization precision and accuracy and outperforms the state of the art.

%\section*{Acknowledgments}
%This work was supported by the European Union's Horizon 2020 research and innovation programme under grant agreement H2020-687786 InVID.

\bibliographystyle{IEEEbib}
\bibliography{refs}

\end{document}